%% file: bspaper_v4_5_sub.tex
\documentclass[prd,twocolumn,showpacs,amsmath,amssymb,superscriptaddress,groupedaddress]{revtex4}
\usepackage{wrapfig,rotating}
\usepackage{amssymb,amsmath,array}
\usepackage{subfigure}
\usepackage{float}
\usepackage{setspace}
\usepackage{graphicx}  % needed for figures
\usepackage{dcolumn}   % needed for some tables
\usepackage{bm}        % for math

\begin {document}

\hspace{5.2in} \mbox{FERMILAB-PUB-13-020-E}

\title{Search for the rare decay $\bm{B_s^0 \to \mu^+ \mu^-}$ \\}

% use the official authorlist for publication
\input author_list.tex

\date{January 18, 2013}

\begin{abstract}

 We perform a search for the rare decay $B_s^0 \to \mu^{+} 
\mu^{-}$ using data collected by the D0 experiment at the 
Fermilab Tevatron Collider.  This result is based on the full D0 Run II 
dataset corresponding to 10.4~fb$^{-1}$ of $p\bar{p}$ collisions 
at $\sqrt{s} = 1.96$ TeV. 
We use a multivariate analysis to increase the sensitivity of the search. 
 In the absence of an observed number of events above the expected 
background, we set an upper limit 
on the decay branching fraction of ${\cal B}$$(B_s^0 \to \mu^{+} 
\mu^{-}) < 15 \times 10^{-9}$  at the 95\% C.L. 
\end{abstract}

\maketitle

\section {Introduction}

\begin {figure}[h]
\begin{center}
\subfigure[]{\includegraphics [width=3.0in] {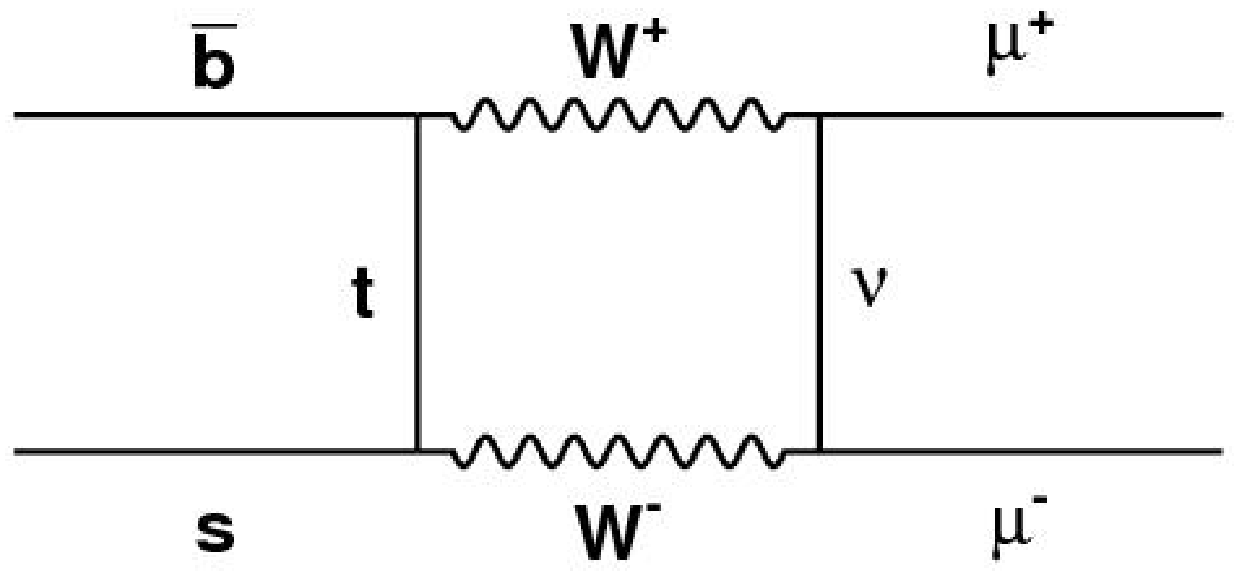}}
\subfigure[]{\includegraphics [width=3.0in] {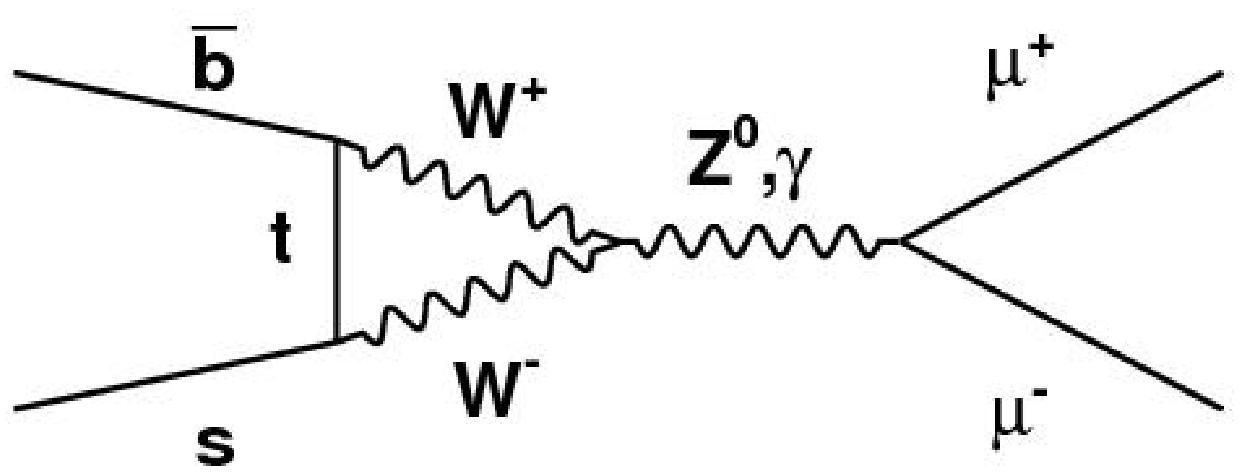}}
\caption{The (a) box diagram and (b) electroweak penguin diagram are examples 
of the FCNC processes through which the decay 
$B_s^0 \to \mu^{+} \mu^{-}$ can proceed.} 
\label{feynman}
\end{center}
\end{figure}

The rare decay $B_s^0 \to \mu^{+} \mu^{-}$ is highly suppressed in the standard 
model (SM) due to its flavor changing neutral current (FCNC) nature. FCNC decays 
can only proceed in the SM through higher-order diagrams as shown in Fig.\ 
\ref{feynman}.  This decay is further suppressed due to the required helicities 
of the final state muons in the decay of the spin zero $B_s^0$ meson. 
 Recent improvements in the SM prediction for the branching fraction 
${\cal B}(B_s^0 \to \mu^+\mu^-)$ include the
effect of the non-zero lifetime difference $\Delta \Gamma_s$ between the
heavy and light mass eigenstates of the $B_s^0$ meson \cite{bruyn,gino}, 
resulting in an expected branching fraction of (3.5$\pm 0.2) \times 10^{-9}$, 
which is about 10\% larger than previous 
calculations \cite{buras}. 

Several scenarios of physics beyond the standard model (BSM) predict significant 
enhancements of this decay channel \cite {bsm1, bsm2, bsm3}, making the study
of this process a
promising way to search for new physics.  However, it is also possible in some 
BSM scenarios for this decay to be suppressed even further than the SM 
prediction \cite{ellis}.

Previous D0 experiment 95\% C.L. limits on the branching fraction for $B_s^0 \to 
\mu^+ \mu^-$ 
include a limit of $ 5 \times 10^{-7}$ from a 
cut-based analysis using 240~pb$^{-1}$ of integrated luminosity 
\cite{d01}; a limit of $ 1.2 \times 10^{-7}$ from a likelihood ratio 
method using an integrated luminosity of 1.3~fb$^{-1}$
 \cite{d02}; and a limit of $5.1 \times 10^{-8}$ using a Bayesian neural network 
and an integrated luminosity of 6.1~fb$^{-1}$ \cite{masato}. 
The result presented here uses the full 
D0 dataset corresponding to 10.4~fb$^{-1}$ of $p\bar{p}$ collisions and 
supersedes our previous results. 

Recently, the LHCb Collaboration has 
presented the first evidence for this decay, at a branching fraction 
consistent 
with the SM prediction \cite{newlhcb}.
 Previous to this result, the most stringent 95\% C.L. limits on this decay 
came from the LHCb 
\cite{lhcb}, CMS \cite{cms}, and ATLAS \cite{atlas} Collaborations, which 
quote 
limits of ${\cal B}(B_s^0 \to \mu^{+} \mu^{-})< 4.5 \times 10^{-9}$, 
$7.7 \times 
10^{-9}$, and $22 \times 10^{-9}$, respectively. 
The CDF Collaboration sees an excess over background corresponding to a 
branching fraction of 
$(18 ^{+11}_{-9}) \times 10^{-9}$ and to a 95\% C.L. upper 
limit of $40 \times 10^{-9}$ \cite {cdf}.

\section{The D0 detector}

The D0 experiment collected data at the Fermilab Tevatron $p\bar{p}$ 
Collider 
at $\sqrt{s}$=1.96 TeV from 2001 through the shutdown of the 
Tevatron in  2011, a period referred to as Run II.  

The D0 detector is described in detail elsewhere \cite{d0det}. For 
the purposes of this analysis, the most important parts of the detector are the 
central tracker and the muon system. The inner region of the D0 central tracker 
consists of a silicon microstrip tracker (SMT) that covers 
pseudorapidities $|\eta|<3$ \cite{eta}. 
In the spring of 
2006, an additional layer of silicon (Layer 0) was added close to the beam pipe 
\cite{layer0}.  Since the detector configuration changed significantly with this 
addition, the D0 dataset is divided into two distinct periods (Run IIa and Run 
IIb), with the analysis performed separately for each period.  Moving away from 
the interaction region, the next detector subsystem encountered is the D0 
central fiber tracker (CFT), which consists of 16 concentric cylinders of 
scintillating fibers, covering $|\eta|<2.5$. Both the SMT 
and CFT are located within a 2 T superconducting solenoidal magnet. The D0 muon 
system is located outside of the finely segmented liquid argon sampling  
calorimeter. The 
muon system consists of three layers of tracking detectors and trigger 
scintillators, one layer in front of 1.8~T toroidal magnets and two 
additional layers after the toroids. The muon system covers $|\eta|< 2$.
 
 The data used in 
this analysis were collected with a suite of single muon and dimuon triggers.  
 
 \section{ Analysis overview} 

This analysis was performed with the relevant dimuon mass region
 blinded until all analysis procedures were final. 
 Our dimuon mass resolution is not
sufficient to separate $B_s^0 \to \mu^+ \mu^-$ from $B_d^0 \to \mu^+ \mu^-$, but
in this analysis we assume that
there is no contribution from $B_d^0 \to \mu^+ \mu^-$, since this decay in expected
to be suppressed with respect to $B_s^0 \to \mu^+ \mu^-$ by the ratio of the CKM
matrix elements $|V_{td}/V_{ts}|^2 \approx 0.04$ \cite{pdg12}. The most stringent
95\%~C.L. limit on the decay  $B_d^0 \to \mu^+ \mu^-$, which is from the LHCb experiment 
\cite{newlhcb}, is ${\cal B}(B^0_d \to \mu^+ \mu^-)<9.4 \times 10^{-10}$.

\begin {figure*}[!th] 
\begin {center} 
\subfigure[]{\label{sig_d}\includegraphics[width=1.7in] {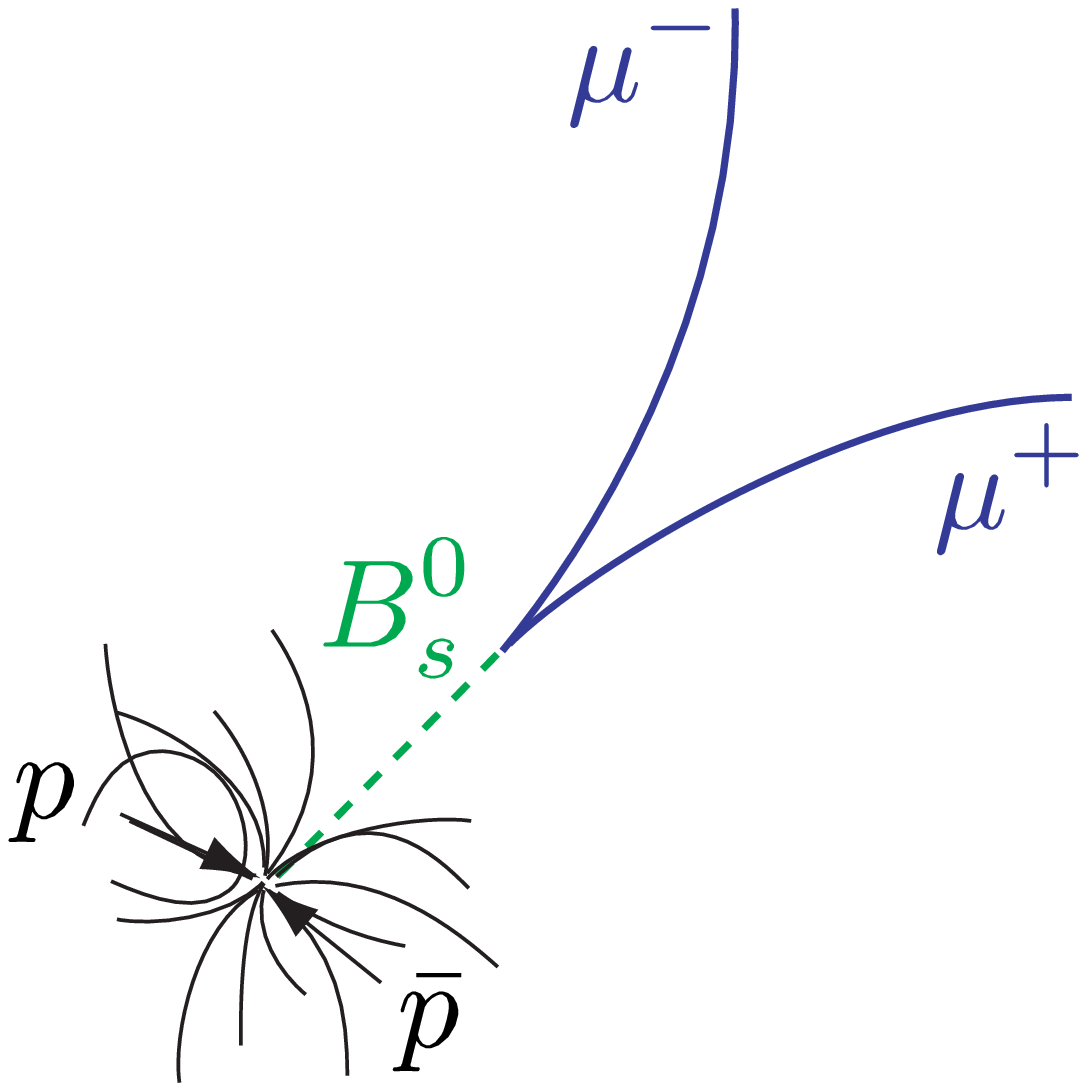}}
\subfigure[]{\label{bg1_d}\includegraphics[width=2.0in] {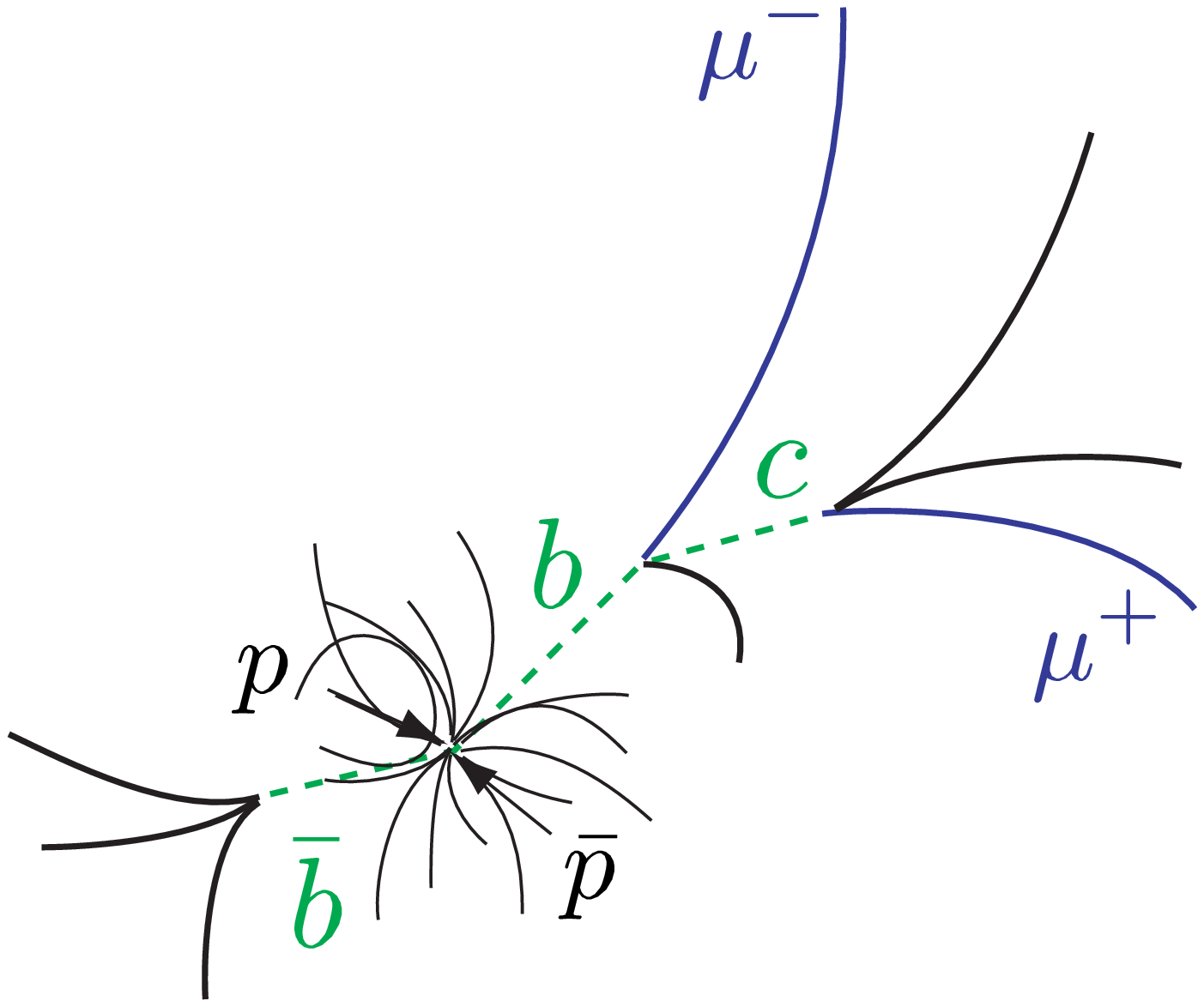}}
\subfigure[]{\label{bg2_d}\includegraphics[width=2.0in] {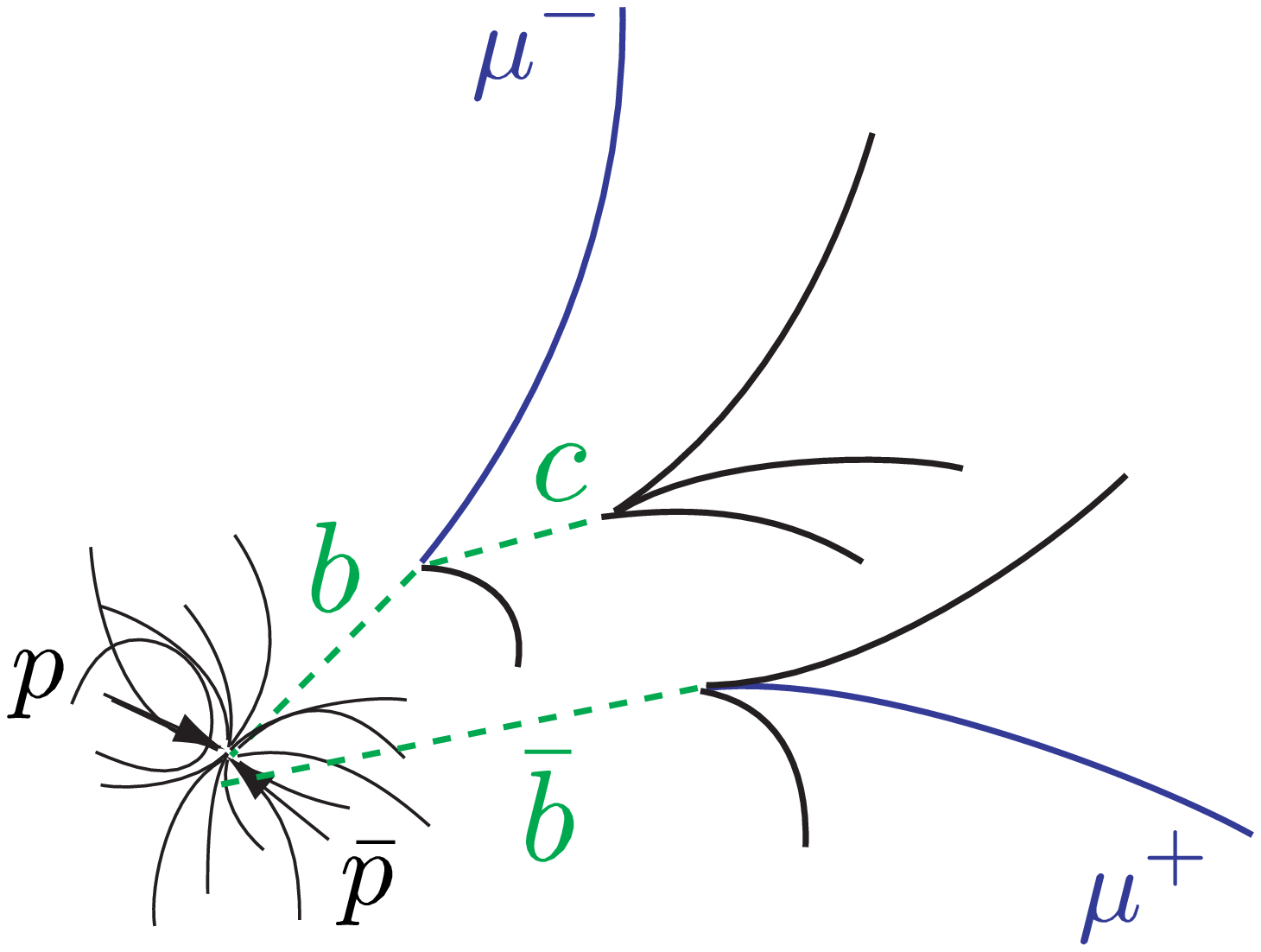}}
\caption {(color online) Schematic diagrams showing (a) the signal decay, 
$B_s^0 \to \mu^{+} \mu^{-}$, 
and main backgrounds:  (b) sequential decay, 
$b \to c \mu^{-}$ followed by $c \to \mu^{+}$, and (c) double semileptonic decay, $b \to \mu^{-}$ and $\bar{b} \to 
\mu^{+}$.} 
\label{cartoon}
\end{center} 
\end{figure*}

 $B_s^0 \to \mu^{+} \mu^{-}$ candidates are identified by selecting two 
high-quality muons of opposite charge that form a good three-dimensional vertex
well-separated from the primary $p\bar{p}$ 
interaction due to the relatively long lifetime of the 
$B_s^0$ meson \cite{pdg12}.  
A crucial requirement 
for this analysis is the suppression of the large dimuon background 
arising from semileptonic $b$ and $c$ quark decays. Figure~\ref{cartoon} 
shows a schematic diagram of the signal decay and the two dominant 
background 
processes. Backgrounds in the dimuon effective mass region below the $B_s^0$ mass 
are dominated by sequential decays such as $b \to \mu^- \nu c $ 
with $c \to \mu^+ \nu X$, as shown in Fig.\ \ref{bg1_d}.  
Backgrounds in the dimuon mass region above the $B_s^0$ mass are 
dominated by double semileptonic decays such as $b(\bar{c}) \to \mu^- \nu X$ and
$\bar{b}(c) \to \mu^+ \nu X$, as
shown in Fig.\ \ref{bg2_d}.
For both of these backgrounds, the muons do not form a real vertex, but the
tracks can occasionally be close enough in space to be reconstructed as 
a ``fake'' vertex. 

Figure \ref{cartoon} illustrates the differences between signal and 
background that we exploit as a general analysis strategy. The dimuon 
system itself should form a good vertex consistent with the decay of a 
single particle originating from the $p\bar{p}$ interaction vertex. 
The $B_s^0$ candidate should have a small impact parameter with respect 
to the primary $p\bar{p}$ interaction vertex, while the individual muons 
should in general have 
fairly large impact parameters. In addition to quantities related to the 
dimuon system, Fig.\ \ref{cartoon} illustrates that the environment 
surrounding the $B_s^0$ candidate should be quite different for signal 
compared to backgrounds. The dimuon system for the signal should be fairly 
well isolated, while the fake dimuon vertex in background events is likely to 
have  
additional tracks and additional vertices nearby. 
No single variable is able to provide definitive discrimination against these 
backgrounds, so we use a multivariate technique as described in  
Sec.~\ref{bdt} to exploit these differences between signal and background.

In addition to dimuon backgrounds from semileptonic heavy quark decays, 
there are peaking backgrounds arising from $B_s^0 \to hh$ or $B_d^0 \to hh$ 
where $hh$ can be $KK$, $K\pi$ or $\pi \pi$. Of these, $B_s^0 \to KK$ is 
the dominant contribution.
 The $K$ or $\pi$ mesons can be misidentified as a muon by decay in flight
$K/\pi \to \mu \nu$
or by penetrating far enough in the detector to create hits in 
the muon system. For these decays to be misidentified as 
signal, both hadrons must be misidentified as a muon, but since the decay 
we are looking for is rare, 
$B_s^0/B_d^0 \to hh$ decays constitute a background of magnitude similar to 
that of the expected signal.   

The number of $B_s^0 \to \mu^+ \mu^-$ decays expected in our dataset is 
determined from analysis of the normalization decay channel $B^{\pm} \to 
J/\psi K^{\pm}$, with $J/\psi \to \mu^+ \mu^-$, as described in detail in 
Sec.~\ref{norm_mode}. 

 \section { Monte Carlo Simulation} \label{mc}

Detailed Monte Carlo (MC) simulations for both the $B_s^0 \to \mu^{+} 
\mu^{-}$ signal and the $B^{\pm} 
\to J/\psi K^{\pm}$ normalization channels are obtained using the {\sc pythia} 
\cite{pythia} event generator, interfaced with the {\sc evtgen} \cite{evtgen} 
decay package. 
The MC includes primary production of $b\bar{b}$ quarks that are
approximately back-to-back in azimuthal angle, and also includes gluon splitting
$g \rightarrow b\bar{b}$ where the gluon may have radiated from any quark
in the event. The latter leads to a relatively collimated $b\bar{b}$ system 
that produces the dominant background when both $b$ and $\bar{b}$ quarks  
decay semileptonically to muons. 

The detector response is simulated using {\sc geant} \cite{geant} and overlaid 
with events from randomly collected $p\bar{p}$ bunch crossings to simulate 
multiple $p\bar{p}$ interactions.  A correction to the MC width of the 
dimuon mass 
distribution is determined from $J/\psi \to \mu^+ \mu^-$ decays in data,  
and this correction is then scaled to the $B_s^0$ mass region. 
The $B_s^0 \to \mu^+ \mu^-$ mass 
distribution in the MC is well described by a double Gaussian function
with the two means 
constrained to be equal, but with the widths ($\sigma_1$ and $\sigma_2$) and 
relative fractions determined by a fit to the corrected mass distribution. 
The average width is 
$\sigma_{av}=f\sigma_1 + (1-f)\sigma_2$=125~MeV, where $f$ is the fraction of the 
area associated with $\sigma_1$. 

\begin {figure*}[!th] 
\begin {center} 
\subfigure[]{\label{ptmu1}\includegraphics [width=3.0in] {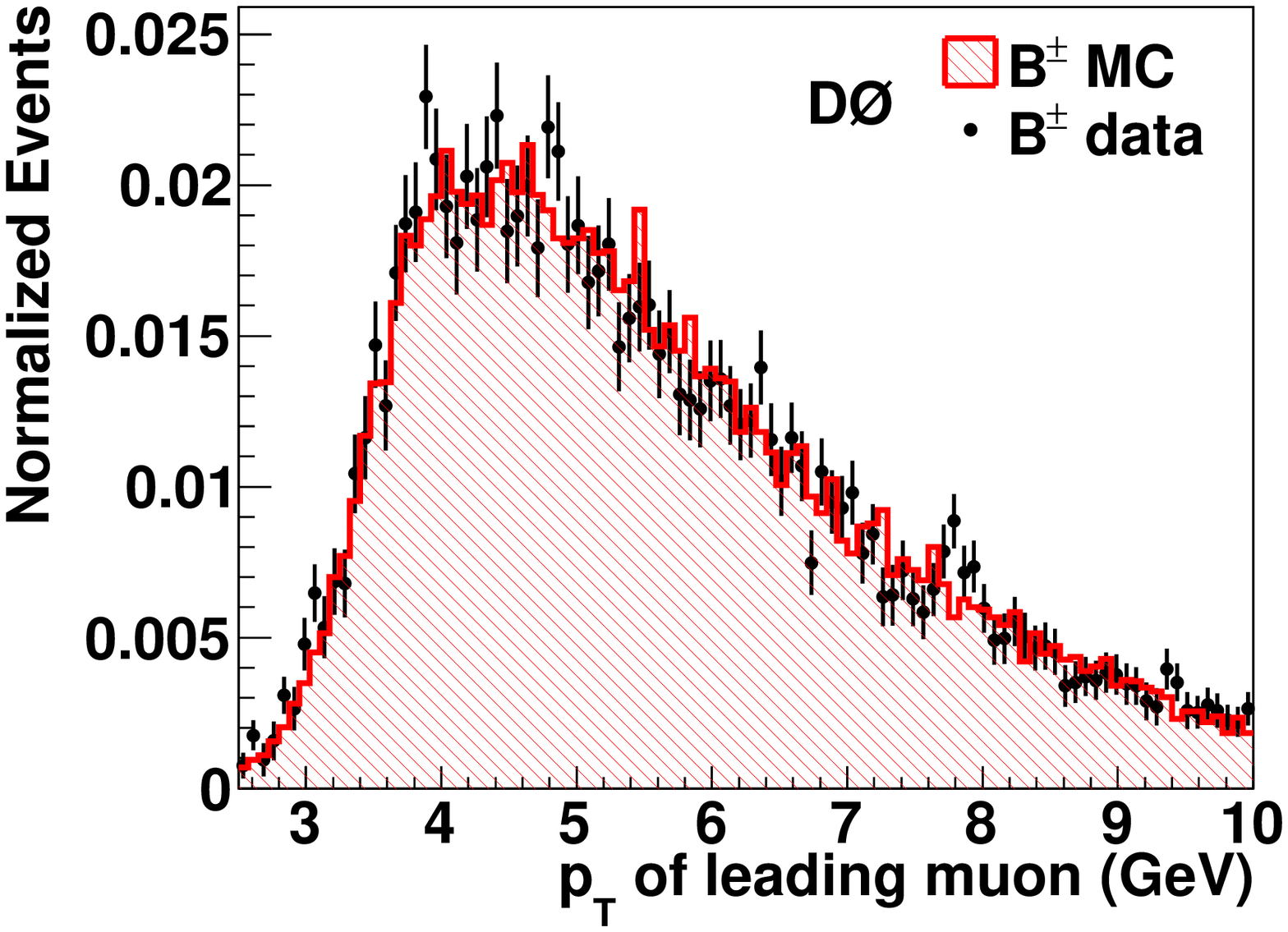}}
\subfigure[]{\label{ptmu2}\includegraphics[width=3.0in] {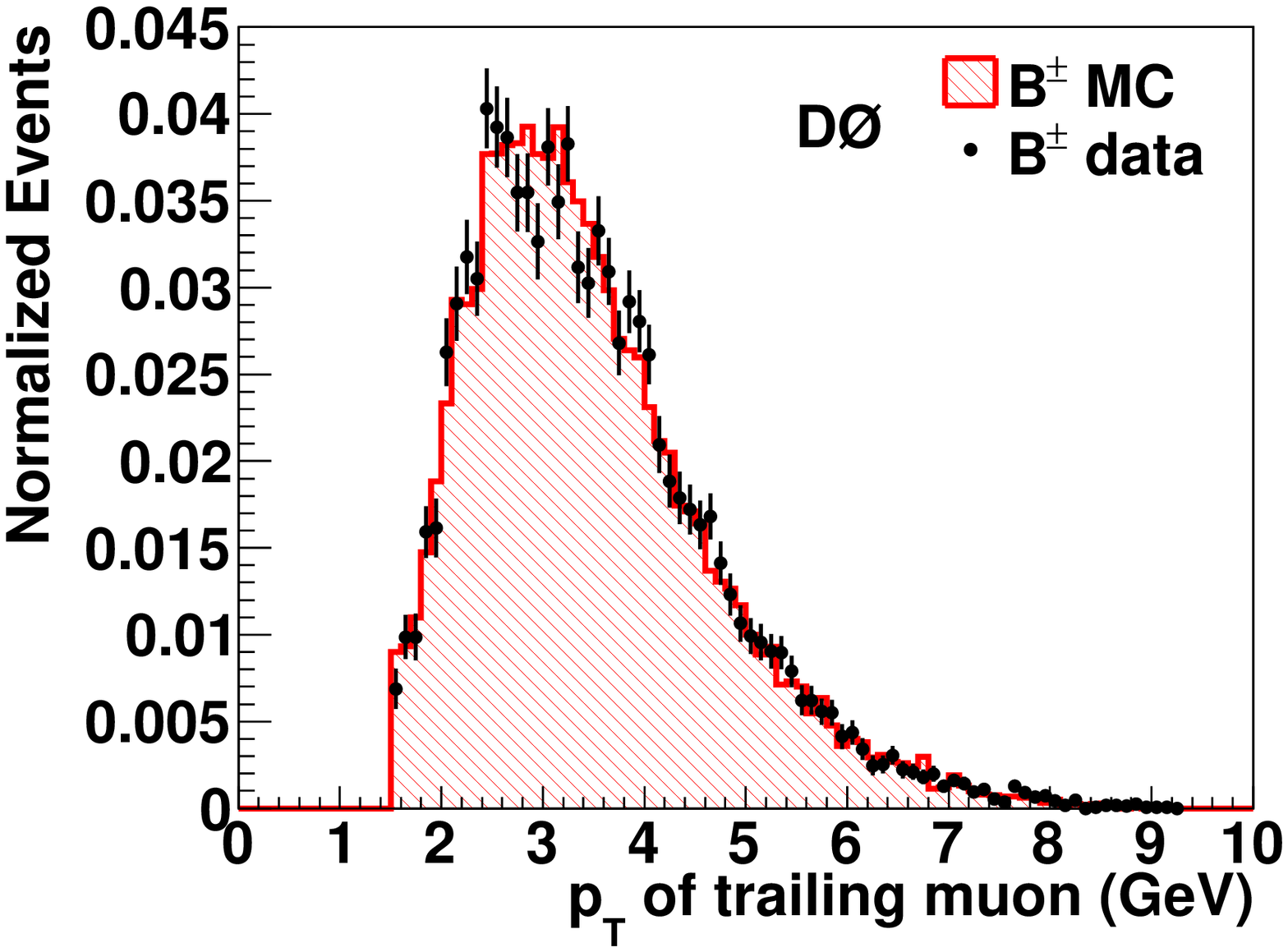}} 
\subfigure[]{\label{ptjpsi}\includegraphics[width=3.0in] {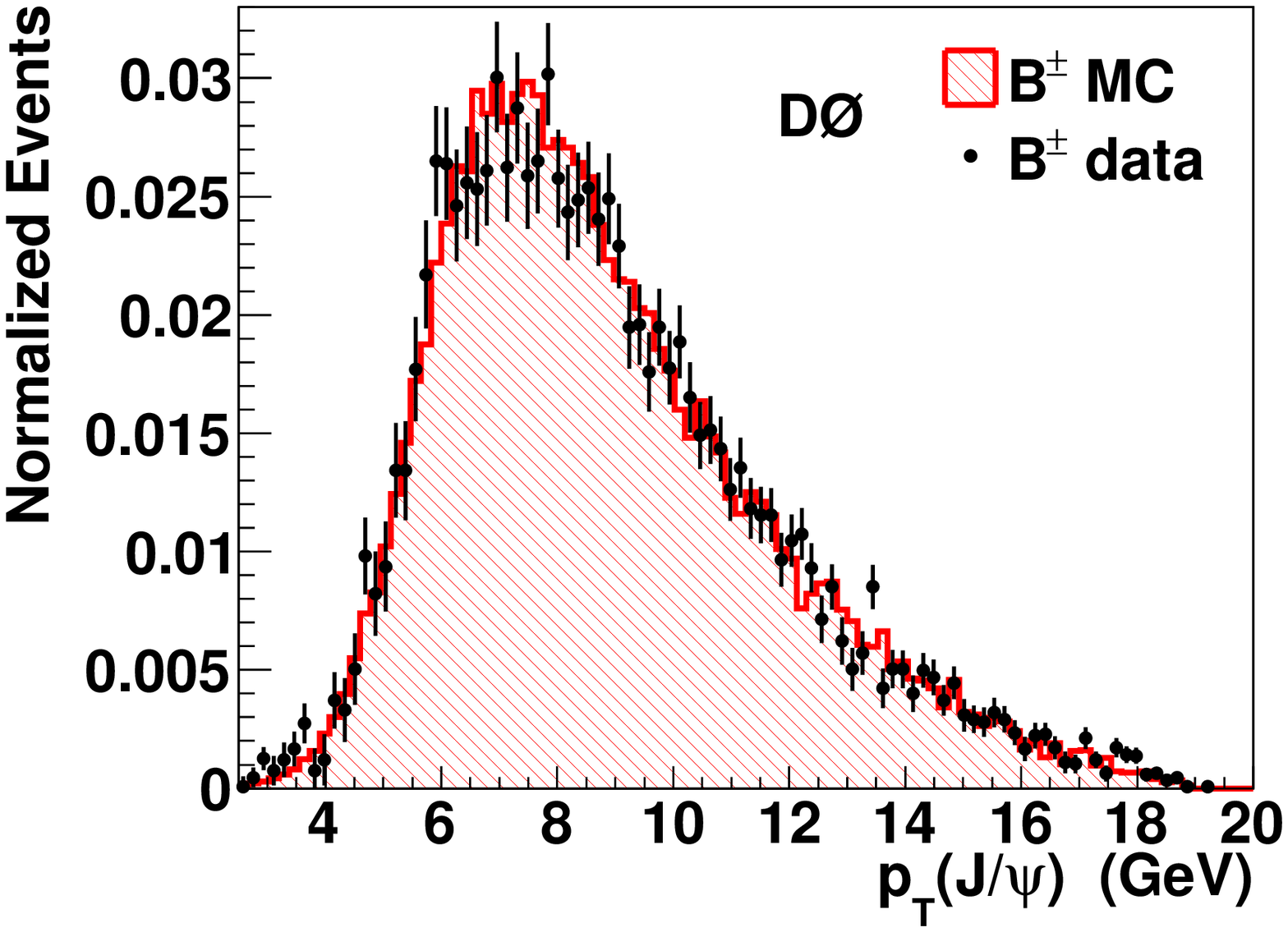}} 
\subfigure[]{\label{Kpt}\includegraphics[width=3.0in] {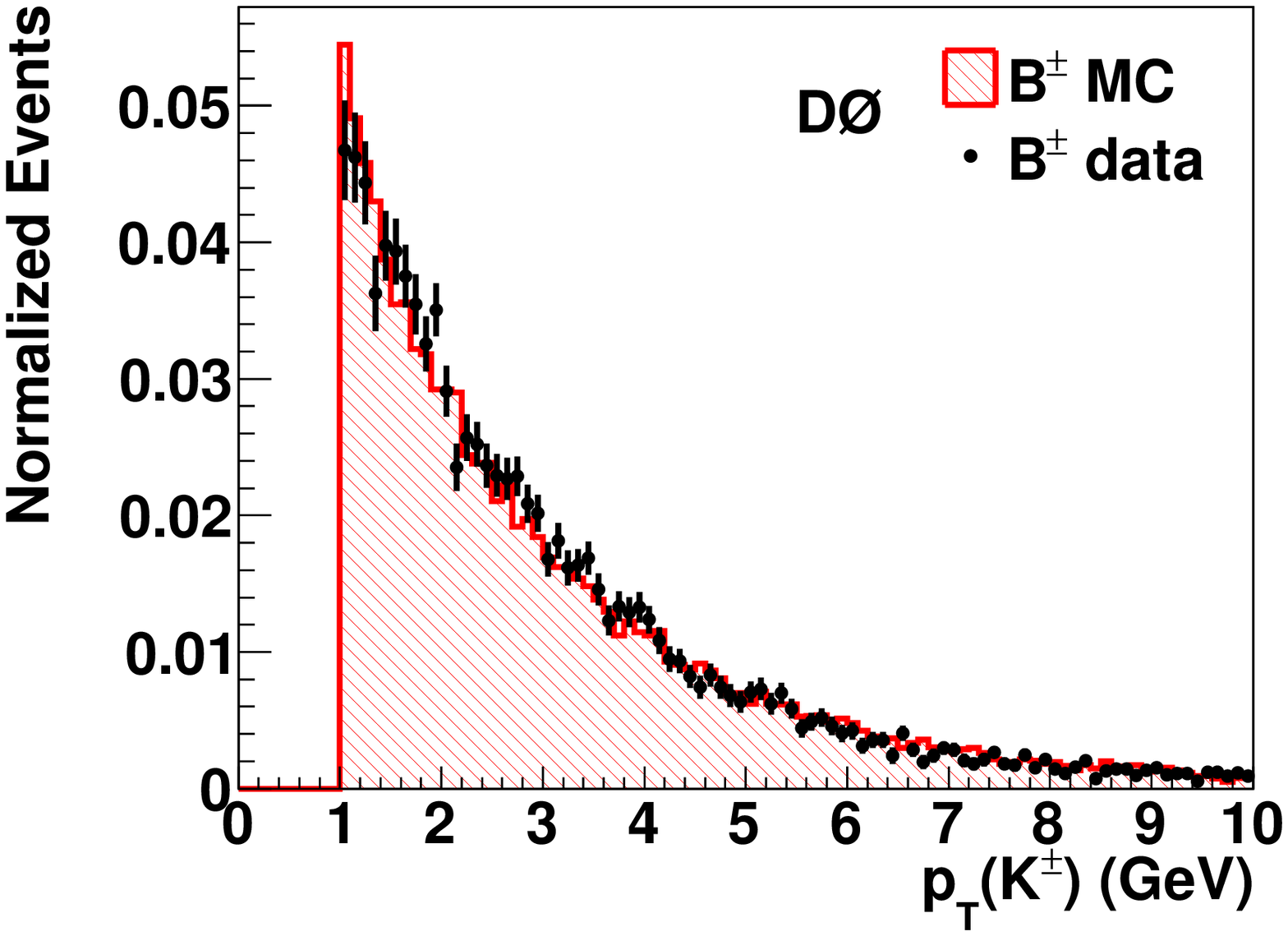}} 
\subfigure[]{\label{Bpt}\includegraphics[width=3.0in] {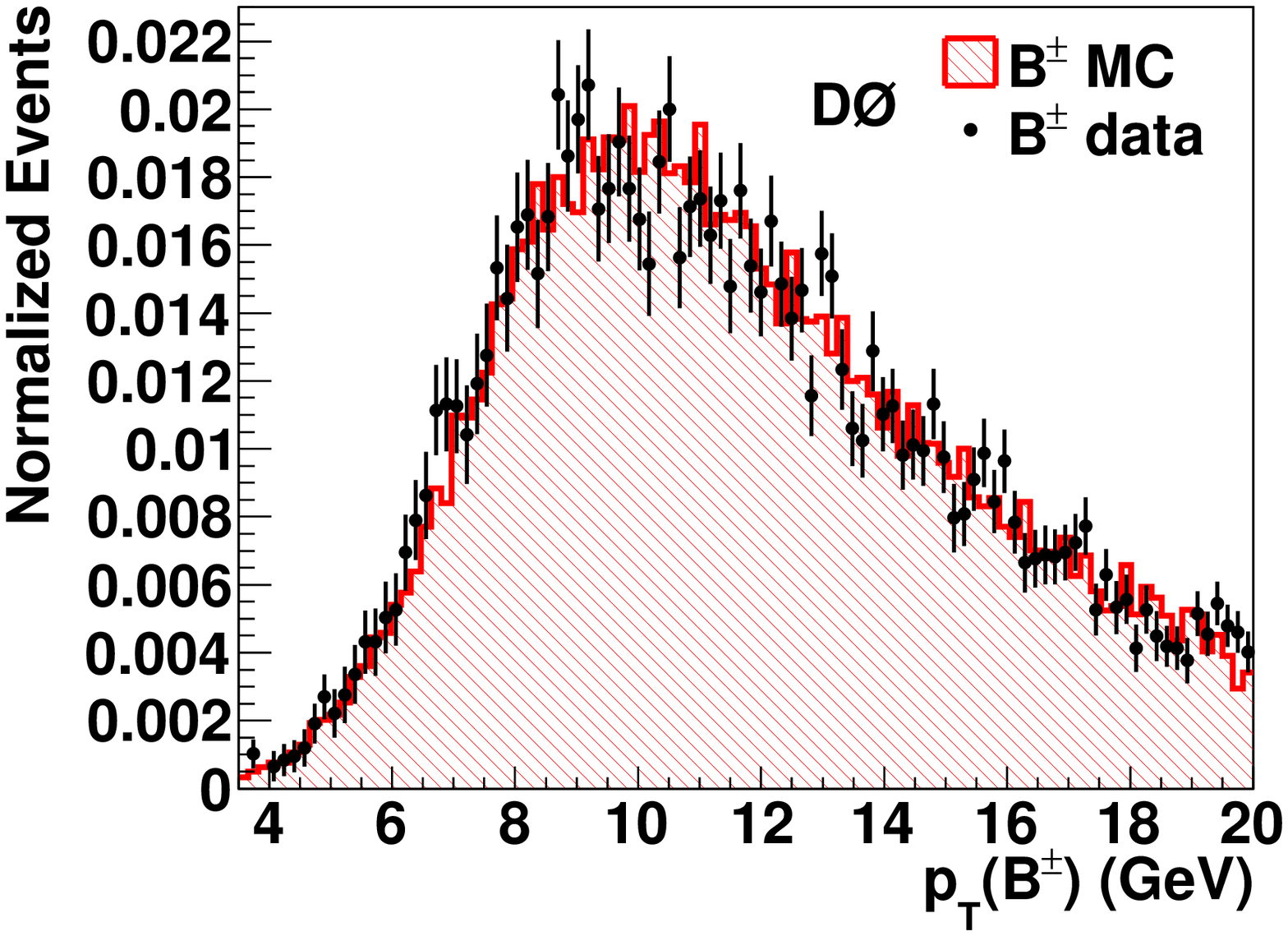}} 
\caption {(color online) Comparison of $p_T$ distributions for data and MC 
simulation, for the normalization channel $B^{\pm} \to J/\psi K^{\pm}$, in 
a single 
data epoch, (a) for the higher-$p_T$ (leading) muon, (b) lower-$p_T$ (trailing) 
muon, (c) $J/\psi$, (d) kaon, and (e) $B^{\pm}$ meson. 
All distributions are normalized to unit area.  } 
\label{pts}
\end{center} 
\end{figure*}

We measure the trigger efficiencies in the data using events 
with no requirements other than a
$p\bar{p}$ bunch crossing (zero-bias events) or events requiring only
an inelastic $p\bar{p}$ interaction 
(minimum-bias events).  
The MC generation does not include trigger efficiencies, 
but the MC events are
reweighted to reproduce the trigger efficiency as a function of the
muon transverse 
momentum ($p_T$).  In addition, the MC events are corrected 
to describe the $p_T$ 
distribution of $B$ mesons above the trigger threshold, as determined from 
$B^{\pm} \to J/\psi K^{\pm}$ decays. Since the trigger conditions changed 
throughout the course of Run II, the $p_T$ corrections are determined separately 
for five different data epochs, with each epoch typically separated by 
an 
accelerator shut-down of a few months' duration.
  Figure~\ref{pts} compares data and MC for several $p_T$ distributions
in the normalization channel, after these corrections.
 The 
background components in the $B^{\pm}$ distributions are removed by 
a side-band subtraction technique, that is, by subtracting the 
corresponding 
distributions from events above and below the 
$B^{\pm}$ mass region. As can be seen 
in Fig.\ \ref{pts}, the $p_T$ distributions in the MC simulation and 
normalization channel data are generally in excellent agreement.
 Figure \ref{pts} shows a single data epoch, but all data epochs show 
similar agreement. 

In addition to the signal MC, we also study the $B_s^0 \to KK$ background 
using a sample 
of MC events that contains about six times the 
expected number of such events in our data 
sample. 

\section { Event selection}

The $B_s^0$ candidate events selected for further study are chosen as follows. 
We select two high-quality, oppositely-charged muons based on information from 
both the central tracker and the muon detectors.  
The primary vertex (PV) of each $p\bar{p}$ interaction is defined  
using all available well-reconstructed tracks and constrained by the mean 
beam-spot position in the transverse plane. 
If a bunch crossing has more than one $p\bar{p}$ interaction vertex, we 
ensure that both muons are consistent with originating from the same PV. 
Tracks 
reconstructed in the central tracker are required to have at least two hits in 
both the SMT and CFT detectors. These tracks are extrapolated to the muon 
system, where they are required to match hits observed in the muon detectors. 
Each muon is required to have transverse momentum $p_T>1.5$~GeV and to have 
pseudorapidity $|\eta|<2$.  Both muons are required to have hits in the muon 
detectors in front of the toroids, and at least one of the muons must also have 
hits in at least one of the muon layers beyond the toroids.  To reduce 
combinatorial backgrounds, the two muons must form a three-dimensional vertex 
with $\chi^2/dof<14$. The dimuon vertex is required to be well separated from 
the PV by examining the transverse decay 
length.  The transverse decay length $L_T$ is defined as $L_T = \vec{l}_T \cdot 
\vec{p}_T/|\vec{p}_T|$, where the vector $\vec{l}_T$ is from the PV
to the dimuon vertex in the transverse plane, and 
$\vec{p}_T$ is the transverse 
momentum vector of the dimuon system.  The quantity  
$\sigma_{L_T}$ is the 
uncertainty on the transverse decay length determined from track parameter 
uncertainties and the uncertainty in the position of the PV.  
To reduce prompt backgrounds, the 
transverse decay length significance of the dimuon vertex, $L_T/\sigma_{L_T}$, 
must be greater than three. Events are selected for further study if the dimuon 
mass $M_{\mu \mu}$ is between 4.0~GeV and 7.0~GeV. These criteria are intended 
to be fairly loose to maintain high signal efficiency, with further 
discrimination provided by the multivariate technique discussed in 
Sec.~\ref{bdt}.

The normalization channel decays $B^{\pm} \to J/\psi K^{\pm}$ with 
$J/\psi \to \mu^+\mu^-$ are 
reconstructed in the data by 
first finding the decay $J/\psi \to \mu^+ \mu^-$ and then adding a third 
track, assumed to be a charged kaon, to the dimuon vertex.  
The selection criteria 
for the signal and normalization channel are kept as similar as possible. 
In 
addition to the above requirements on the muons, we require the $K^{\pm}$ 
to have $p_T >$ 1 GeV and $|\eta|<2$, and we require the three-track vertex to 
have $\chi^2/dof<6.7$. In the normalization channel the dimuon mass is required 
to be in 
the $J/\psi$ mass region, 2.7~GeV $<M(\mu^+\mu^-)<$ 3.45~GeV.

 \section{ Determination of the Single Event Sensitivity}\label{norm_mode} 

To determine the number of $B_s^0 \to \mu^+ \mu^-$ decays we expect in the data, we 
normalize to the number of $B^{\pm}\to J/\psi K^{\pm}$ candidates observed in the data. 
The number of $B^{\pm}\to J/\psi K^{\pm}$ decays is used to determine the 
single event sensitivity (SES), defined as the branching fraction 
for which one event is expected to be present in the dataset. The SES
is calculated from \\

\hspace*{.2in}SES $=\frac{1}{N(B^{\pm})} \times 
\frac{\epsilon(B^{\pm})}{\epsilon(B_s^0)} \frac{f(b \to B^{\pm})}{f(b\to B_s^0)} \times $ 
\\
\hspace*{.8in}${\cal B}(B^{\pm} \to J/\psi K^{\pm})\times $$\cal{B}$$(J/\psi \to 
\mu^{+} \mu^{-})$. \\
\\
In this expression $N(B^{\pm})$ is the number of $B^{\pm} \to J/\psi K^{\pm}$ 
decays observed in the data, as discussed below. 
The efficiency for reconstructing the 
normalization channel decay, $\epsilon(B^{\pm})$, and the signal channel,
$\epsilon(B_s^0)$, are determined from MC 
simulations as discussed in more detail below. The fragmentation 
ratio $f(b\to B^{\pm})/f(b \to B_s^0)$ is the 
relative probability of a $b$ quark fragmenting to a $B^{\pm}$ compared to a 
$B_s^0$. We use the 
``high energy'' average $f(b\to B_s^0)/f(b\to 
B^{\pm})$ = 0.263 $\pm$ 0.017 
provided by the Heavy Flavor Averaging Group \cite{hfag} for the 
2012 Particle Data Group compilation \cite{pdg12}, which is 
consistent with other recent measurements \cite{frag}.  The product of the 
branching 
fractions $\cal{B}$$(B^{\pm} \to J/\psi K^{\pm})   
\times $$\cal{B}$$(J/\psi \to \mu^{+} \mu^{-})$ is $(6.01 \pm 0.21) \times 
10^{-5}$ \cite{pdg12}.  

\begin {figure} [h] 
\begin{center} 
\includegraphics [width=3.5in] {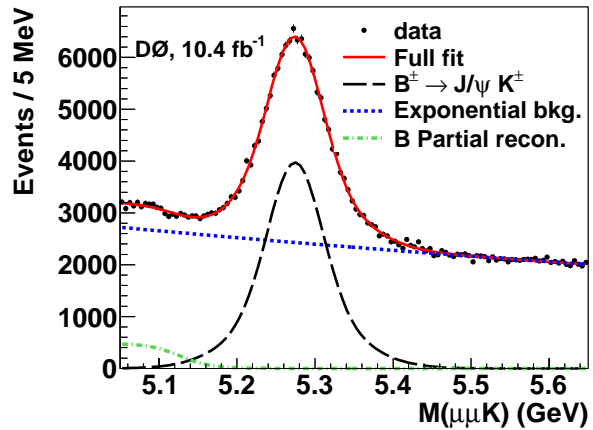} 
\caption{ (color online) Invariant mass distribution for the normalization 
channel 
$B^{\pm} \to J/\psi K^{\pm}$ for the entire Run II dataset. The 
full fit is shown as the solid line, the $B^{\pm} \to J/\psi K^{\pm}$ 
contribution is shown as the dashed line, the exponential background is 
shown as
the dotted line, and the contribution from partially reconstructed $B$ 
meson decays is shown as the dot-dash line. } 
\label{norm} 
\end{center} 
\end{figure}

Figure~\ref{norm} 
shows the normalization channel mass distribution, $M(\mu^+ \mu^- K)$, for the 
entire Run II dataset.
The mass distribution is fitted to a 
double Gaussian function to model the normalization channel decay and an exponential 
function to model the dominant background. A hyperbolic tangent 
threshold function is also included in the fit to 
model partially reconstructed $B$ meson decays, primarily $B^0_d \to J/\psi 
K^{0*}$.  A possible contribution from  
$B^{\pm} \to J/\psi \pi^{\pm}$ is also included in the fit, although this 
contribution is not statistically significant and is not shown 
in the Fig.~\ref{norm}. Systematic uncertainties on $N(B^{\pm})$ are determined 
from 
variations in the mass range of the fit, the histogram binning, and the 
background model. An additional systematic uncertaintity on $N(B^{\pm})$ is due to the 
candidate selection.  If an event has more than one $B^{\pm} \to J/\psi K^{\pm}$
candidate, we retain only the candidate with the best vertex $\chi^2$. This 
choice results in fewer overall reconstructed $B^{\pm} \to J/\psi K^{\pm}$
decays but also less background. To determine the systematic effect due to this 
choice, we have reconstructed $B^{\pm} \to J/\psi K^{\pm}$ decays in two of the 
five data epochs retaining all candidates. The SES depends on the ratio 
$N(B^{\pm})/\epsilon(B^{\pm})$, and we find that this ratio varies at most 
2.2\%, which we take 
as an additional systematic uncertainty on $N(B^{\pm})$. 
 We observe a total of 
 $(87.4\pm 3.0)\times 10^3$ $B^{\pm} \to J/\psi K^{\pm}$ decays in the full 
dataset, where the uncertainty includes both statistical and 
systematic effects.

The ratio of reconstruction efficiencies that enters into the SES is 
determined 
from MC simulation. 
One source  of systematic uncertainty in the efficiency ratio arises from 
the trigger efficiency corrections applied to the MC, as described in 
Sec.~\ref{mc}. The variation 
in these corrections over data epochs with similar trigger conditions 
allows us
to set a  1.5\% systematic uncertainty on the efficiency ratio due to this 
source. An additional systematic uncertainty arises from the 
 efficiency for finding a third track. There could be a data/MC 
discrepancy in this efficiency which will not cancel in the ratio. 
We evaluate this systematic uncertainty by 
comparing the efficiency for finding an extra track in 
data and MC in the four-track decay $B^0_d \to J/\psi K^{0*}$ with $K^{0*} \to 
K\pi$ and in the three-track normalization channel decay $B^{\pm} \to J/\psi K^{\pm}$. 
From this study, we determine that the data/MC efficiency ratio for 
identifying the third track varies with data epoch but is on average 0.88 $\pm$ 
0.06, where the uncertainty 
includes statistical uncertainties from the fits used to extract the 
number of signal 
events, and systematic uncertainties estimated from fit variations.  
The efficiency for $B^{\pm}$ reconstruction is adjusted in each 
data epoch for this track-finding efficiency correction.  
The reconstruction efficiency ratio $\epsilon(B^{\pm})/\epsilon(B_s^0)$ is 
determined to be (13.0 $\pm$ 0.5)\% on average, but 
varies over the different data epochs by about 1.0\%.  
The efficiency for the  $B^{\pm} \to J/\psi K^{\pm}$ decay is impacted by the softer
$p_T$ distribution of the muons in the three-body decay as well as the 
fairly hard ($p_T>1$~GeV) cut on the $p_T$ of the kaon, and the candidate selection which 
retains only the three-track candidate with the best vertex $\chi^2$. 
 
When all statistical and systematic uncertainties are taken into account, 
the SES is found to be $(0.336 \pm 0.029) \times 10^{-9}$ 
before the multivariate selection, yielding a SM expected number of $B_s^0 \to 
\mu^{+} \mu^{-}$ events of 10.4 $\pm$ 1.1 events in our data sample.

 \section {Multivariate Discriminant} \label{bdt}

 A boosted decision tree (BDT) algorithm, as implemented in the {\sc tmva} 
package 
of  {\sc ROOT} \cite{tmva}, is used to differentiate between signal 
and the 
dominant backgrounds. The BDT is trained using MC simulation for 
the signal and data sidebands for the background. The data sidebands include events 
in the dimuon mass range 4.0--4.9~GeV (low-mass sidebands)  and 5.8--7.0~GeV (high-mass 
sidebands), with all selection cuts applied. 
The low-mass sidebands 
are dominated by sequential decays, illustrated in Fig.~\ref{bg1_d}, 
while the the high-mass sidebands are dominated by 
double $B$ hadron decays, as illustrated in Fig.~\ref{bg2_d}. We therefore 
train  two BDTs  
to separately discriminate against these two backgrounds.
Each BDT discriminant uses 30 variables that 
fall into two general classes.  

\begin {figure*} [!th]
\begin{center}
\subfigure[]{\label{ip1} \includegraphics [width=3.0in] {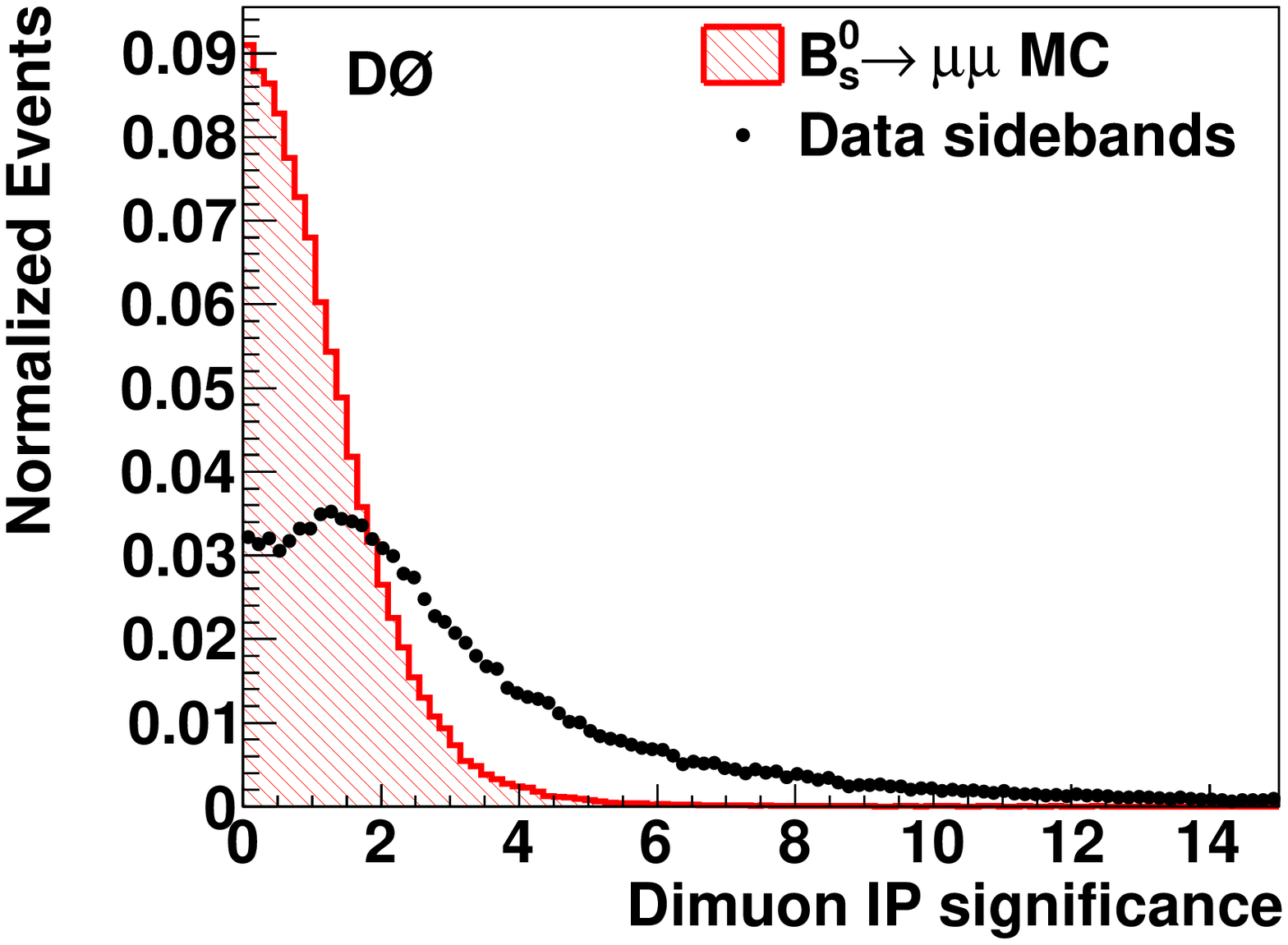}}
\subfigure[]{\label {ip2}\includegraphics [width=3.0in] {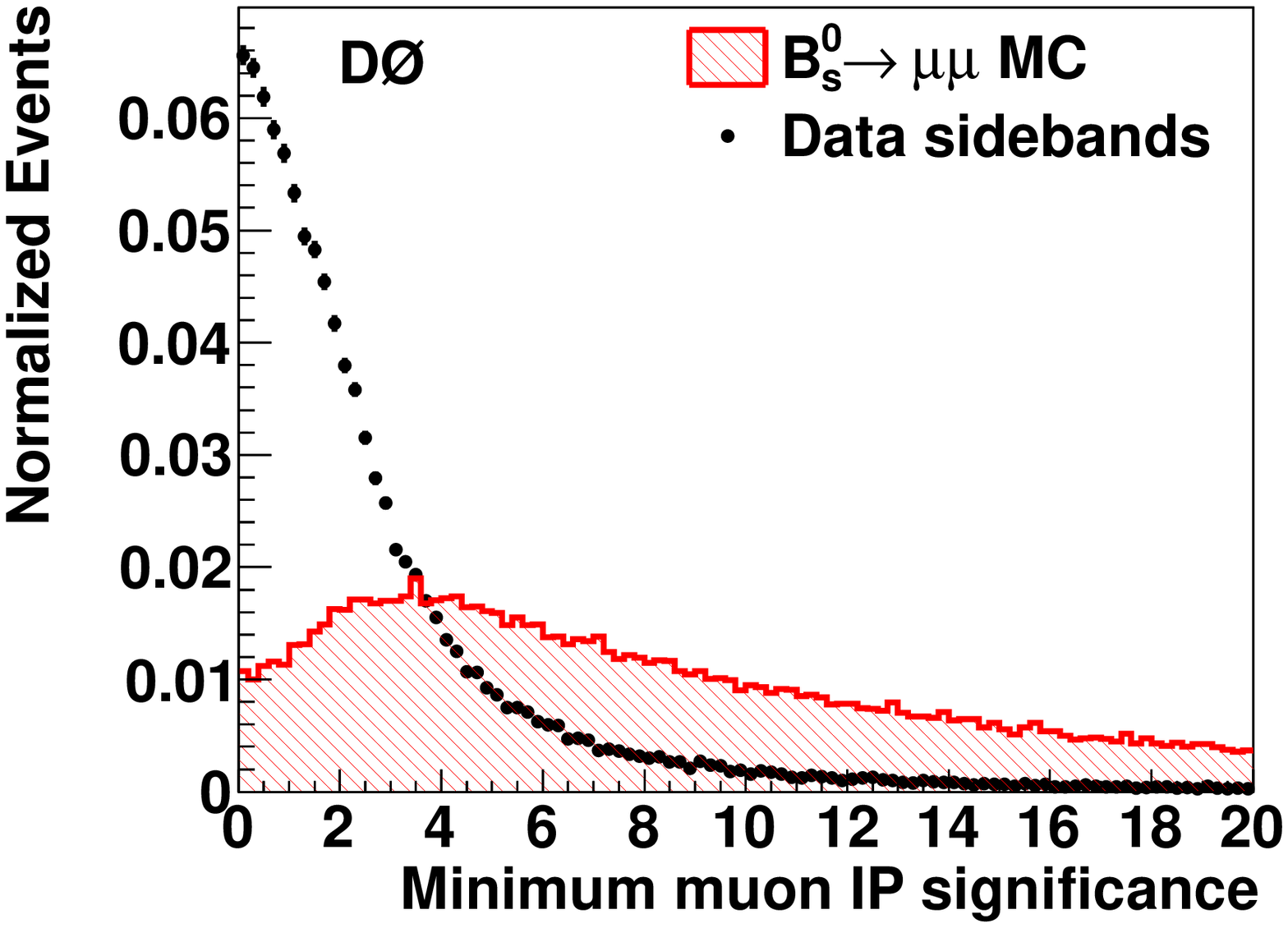}}
\caption{(color online) Comparison of signal MC and background sideband 
data for (a) the $B_s^0$ candidate impact parameter significance and (b) 
the minimum muon impact parameter significance.
All distributions are normalized to unit area. }
\label{ip}
\end{center}
\end{figure*}

One class of variables includes kinematic and topological quantities 
related to the dimuon system. These variables include the pointing angle, 
defined as the angle between the dimuon momentum vector $\vec{p}(\mu^+\mu^-)$ 
and the vector from the PV to the dimuon vertex. 
The dimuon $p_T$ 
and impact parameter, as well as the $p_T$ values of the individual muons 
and their impact
parameters, are also used as discriminating variables. As examples of dimuon 
system variables that discriminate between signal and background, Fig.\ 
\ref{ip1} shows the impact parameter significance (impact parameter divided by 
its uncertainty) of the $B_s^0$ candidate for signal MC and background, and 
Fig.\ \ref{ip2} shows the minimum impact parameter significance for 
the individual muons, that is, the smaller of the two values.

\begin {figure*} [!th]
\begin{center}
\subfigure[]{\label{iso1} \includegraphics [width=3.0in] {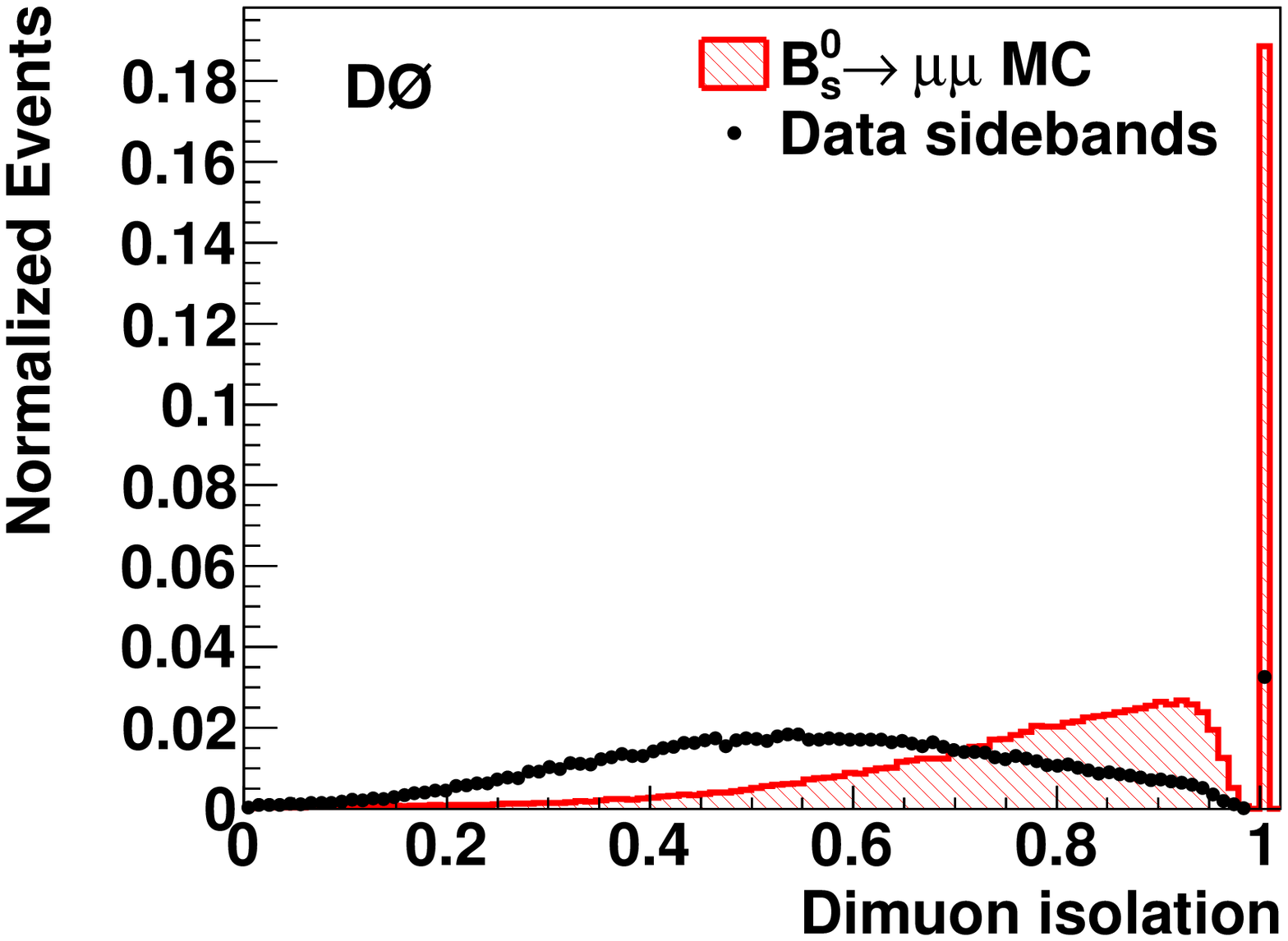}}
\subfigure[]{\label {iso2}\includegraphics [width=3.0in] {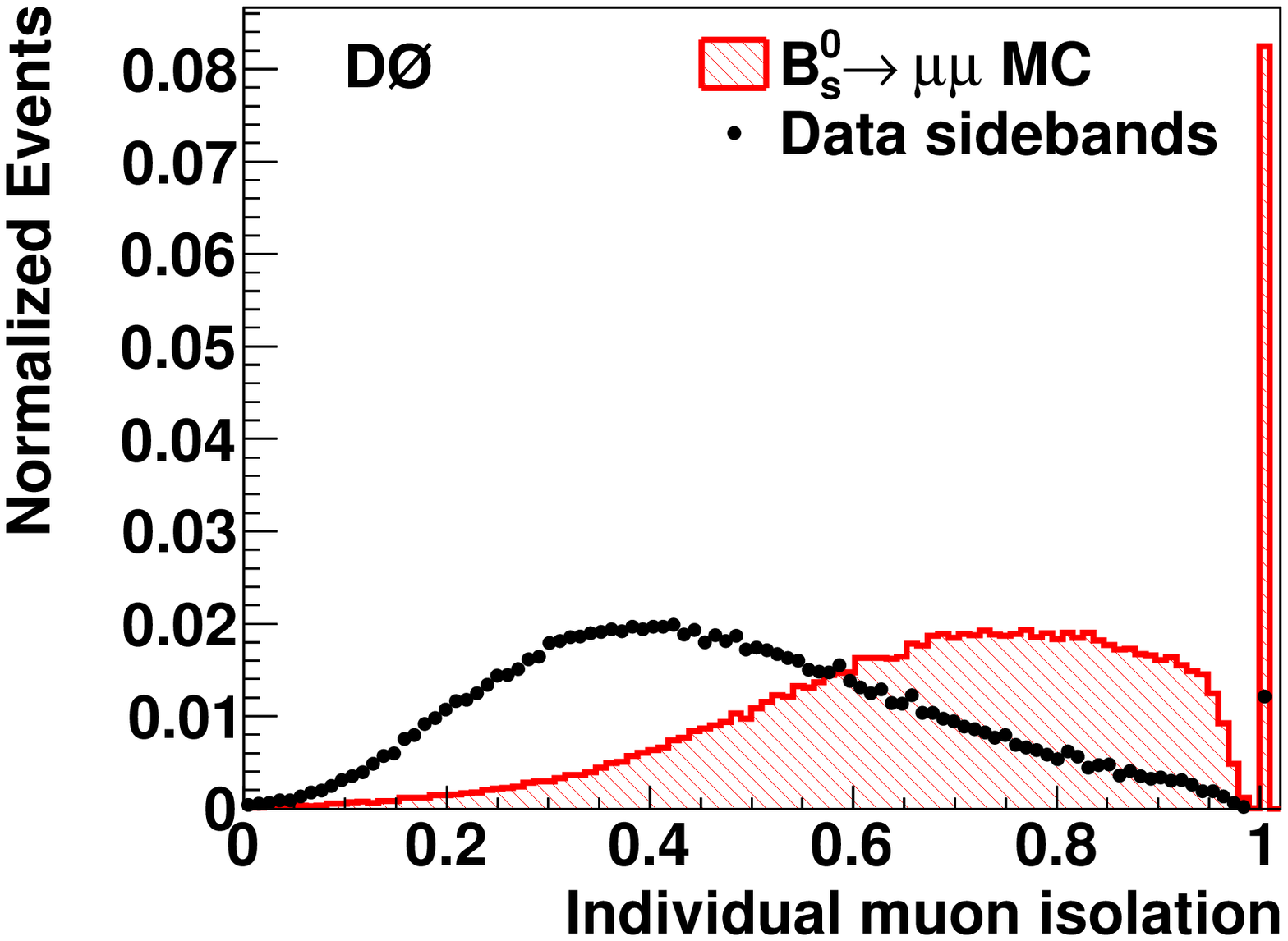}}
\caption{(color online) Comparison of signal MC and background sideband 
data for (a) isolation defined with respect to the dimuon system and (b) for the average 
of the two isolations defined with respect to the individual muons.
All distributions are normalized to unit area. }
\label{isolation}
\end{center}
\end{figure*}

A second general class of variables used in the BDT discriminants includes 
various isolation-related quantities. Isolation is defined with respect to a 
momentum vector $\vec{p}$ by constructing a cone in azimuthal angle $\phi$ 
and pseudorapidity $\eta$ around the momentum vector, with the cone radius defined by 
${\cal R} = \sqrt{\Delta \eta^2 + \Delta \phi^2}$. The isolation ${\cal I}$ 
is then 
defined as ${\cal I} = p_T/[p_T + p_T(\text{cone}])$
 where $p_T(\text{cone})$ is the scalar sum of the $p_T$ of all tracks 
(excluding the 
track of interest) with $\cal{R}$ less  than some cut-off value, chosen to 
be ${\cal R}=1$ in 
this analysis.  For a perfectly isolated track (that is, no other tracks in the 
 cone), ${\cal I} = 1$.  Figure \ref{cartoon} shows that background events 
are expected to be 
less isolated than signal events. For maximum signal/background discrimination, we 
define isolation cones around the dimuon direction and around each muon 
individually. From simulation studies, we find that for background events, the 
two muons are often fairly well separated in space, so using individual 
isolation 
cones around each muon adds discriminating power. Figure~\ref{isolation} compares 
signal MC and data sidebands for two examples of isolation variables.

We also search for additional vertices near the dimuon vertex using 
two different techniques.  As illustrated by Fig.\ \ref{cartoon}, in 
background events the muons often form a good vertex with another charged 
track. We try to reconstruct such vertices using tracks that are 
associated with the same PV as the dimuon pair, which have an 
impact parameter with respect to the PV of at least 30 microns, 
and which have an impact parameter significance of at least 3.0. 
If a track satisfying these requirements 
forms a vertex with one of the muons with a vertex $\chi^2/dof<5.0$, we consider 
this an additional vertex. Additional tracks, satisfying the same
requirements as above,  can be included in this vertex 
if they do not increase the vertex $\chi^2$ by more than 5.0.  
This procedure is carried out with both muons, allowing for the 
possibility of 
finding an additional vertex with either or both of the muons. We also 
attempt to reconstruct additional vertices using tracks that have an impact 
parameter significance with respect to the dimuon vertex of less than 
4.0. We 
allow these vertices to include or not include one of the muons. When an additional
vertex is successfully reconstructed, the vertex $\chi^2$, the 
invariant mass of the particles included in the vertex, and the vertex 
pointing angle are used as discriminating variables in the BDTs. In the 
case where no 
such vertices are found, these variables are set to nonphysical values. We 
find that, for the background sidebands, at least one additional 
vertex is reconstructed 80\% of the time, while for the signal MC, one or more
additional vertices are found 40\% of the time.

\begin {figure*} [th!]
\begin{center}
\subfigure[]{\label{ipnorm1}\includegraphics [width=3.0in] {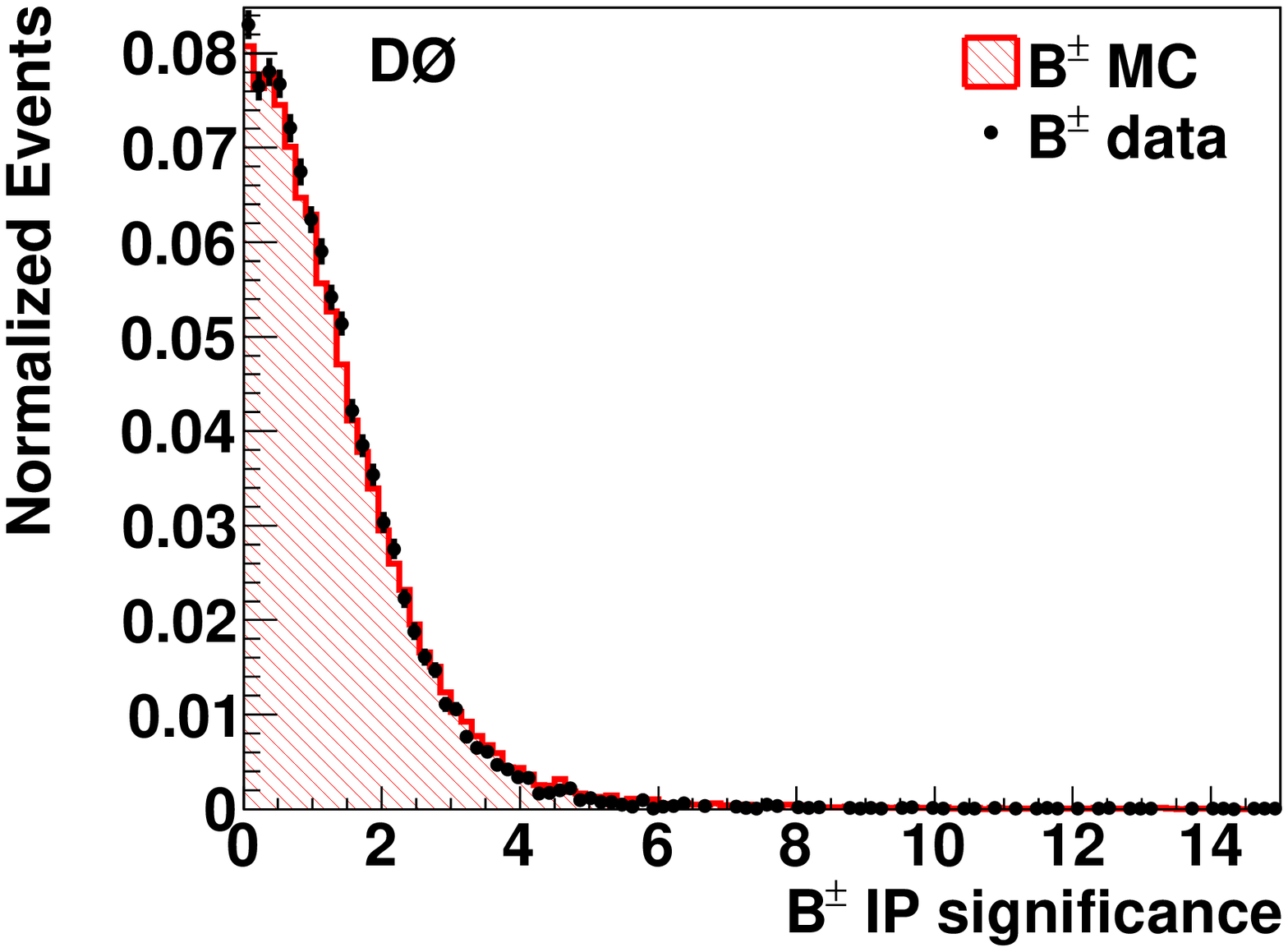}}
\subfigure[]{\label{ipnorm2}\includegraphics [width=3.0in] {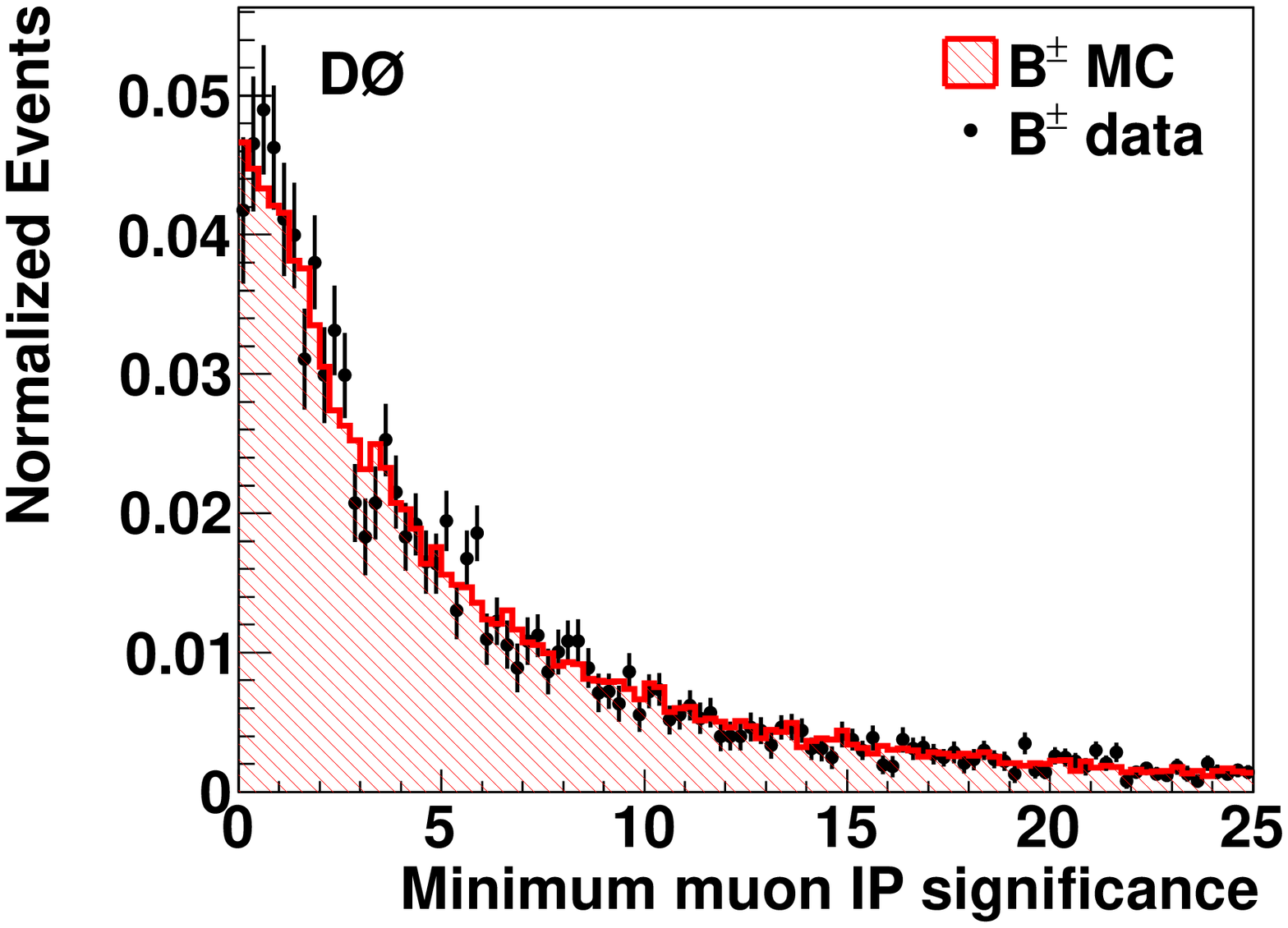}}
\caption{(color online) Comparison of normalization channel MC and 
sideband-subtracted data for (a)
$B^{\pm}$ impact parameter significance and (b) the minimum muon impact 
parameter 
significance. All distributions are normalized to unit area. }
\label{ipsignorm}
\end{center}
\end{figure*}

\begin {figure*} [th!]
\begin{center}
\subfigure[]{\label{isonorm1}\includegraphics [width=3.0in] {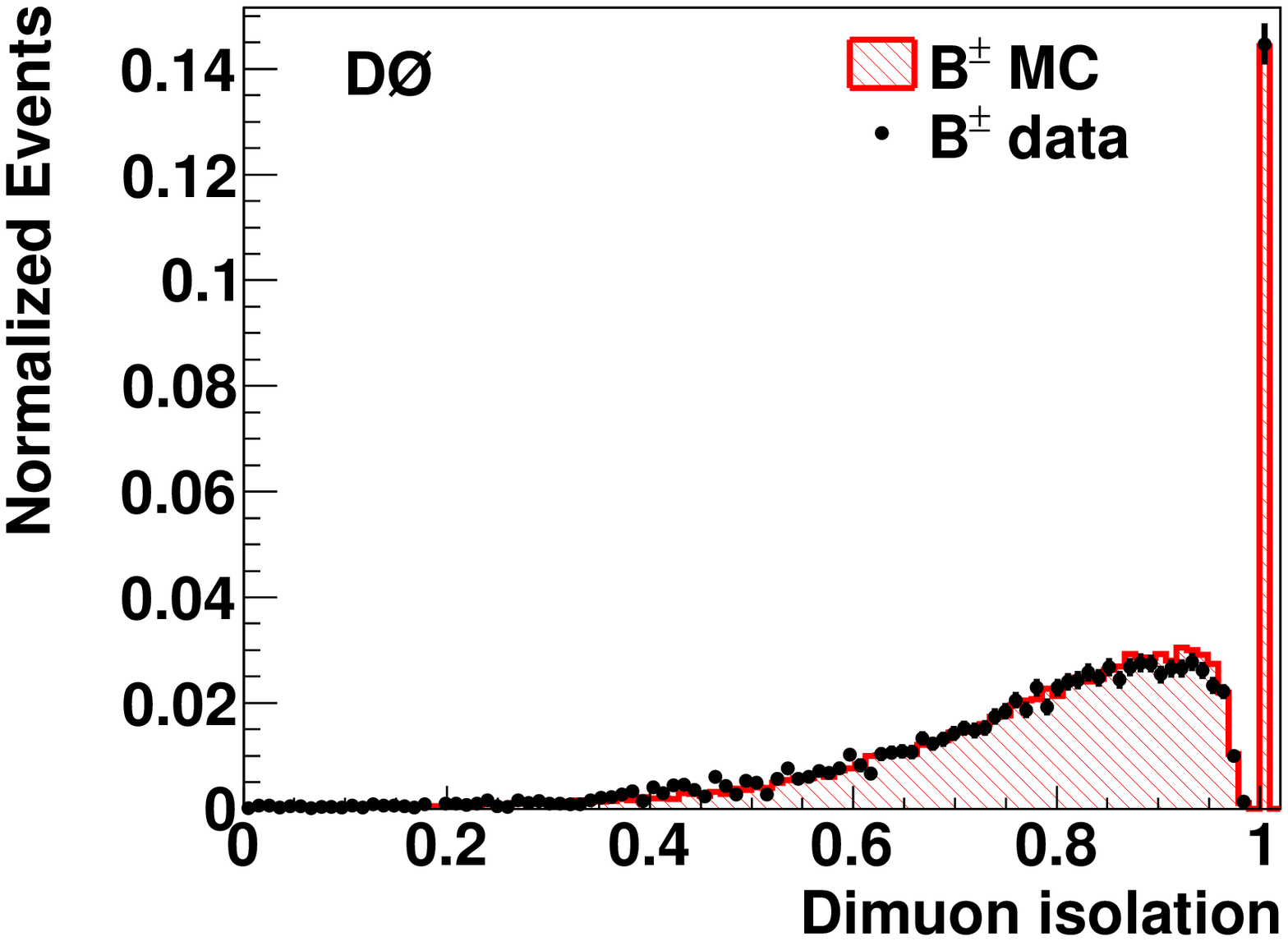}}
\subfigure[]{\label{isonorm2}\includegraphics [width=3.0in] {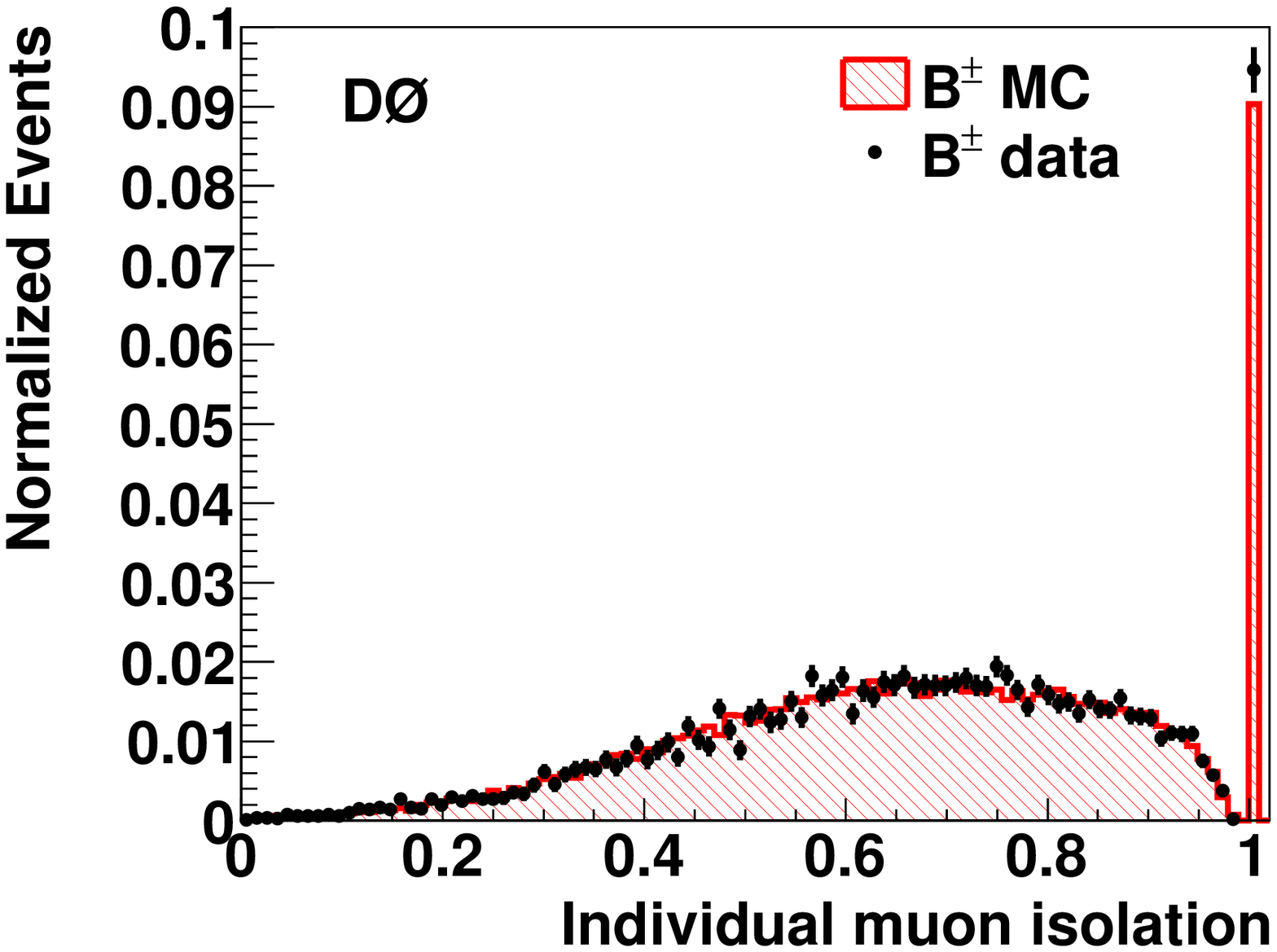}}
\caption{(color online) Comparison of normalization channel MC and 
sideband-subtracted data for 
(a) dimuon isolation and (b) the average of the two individual muon isolations. 
All distributions are normalized to unit area. }
\label{isonorm}
\end{center}
\end{figure*}

To verify that the MC simulation is a good representation of the data, 
 we compare the sideband-subtracted normalization channel data with the 
normalization channel MC. Figure~\ref{ipsignorm} compares the normalization 
channel data and the MC simulation for the $B^{\pm}$ meson impact 
parameter 
significance 
and the minimum muon impact parameter significance. Figure~\ref{isonorm} 
shows the same comparison for the dimuon and individual muon isolation 
variables.
 We check all 30 variables used in the multivariate 
discriminant to confirm good agreement between data and MC for the 
normalization channel.

We make additional requirements on both the data sidebands and the signal MC 
before events are used in the BDT training. These requirements include dimuon 
$p_T >5$~GeV and the cosine of the dimuon pointing angle $>0.95$.  These 
requirements are 78\% efficient on average in retaining signal events but exclude about 
96\% of the background. We find a significant enhancement in background 
rejection from the BDT discriminants using these additional requirements before 
BDT training. These requirements are (93 $\pm$ 1)\% efficient for the normalization 
mode MC, and (91 $\pm$ 3) \% efficient for the  normalization mode data. 

\begin {figure*} [ht]
\begin{center}
\subfigure[]{\label{KS1}\includegraphics [width=3.0in] {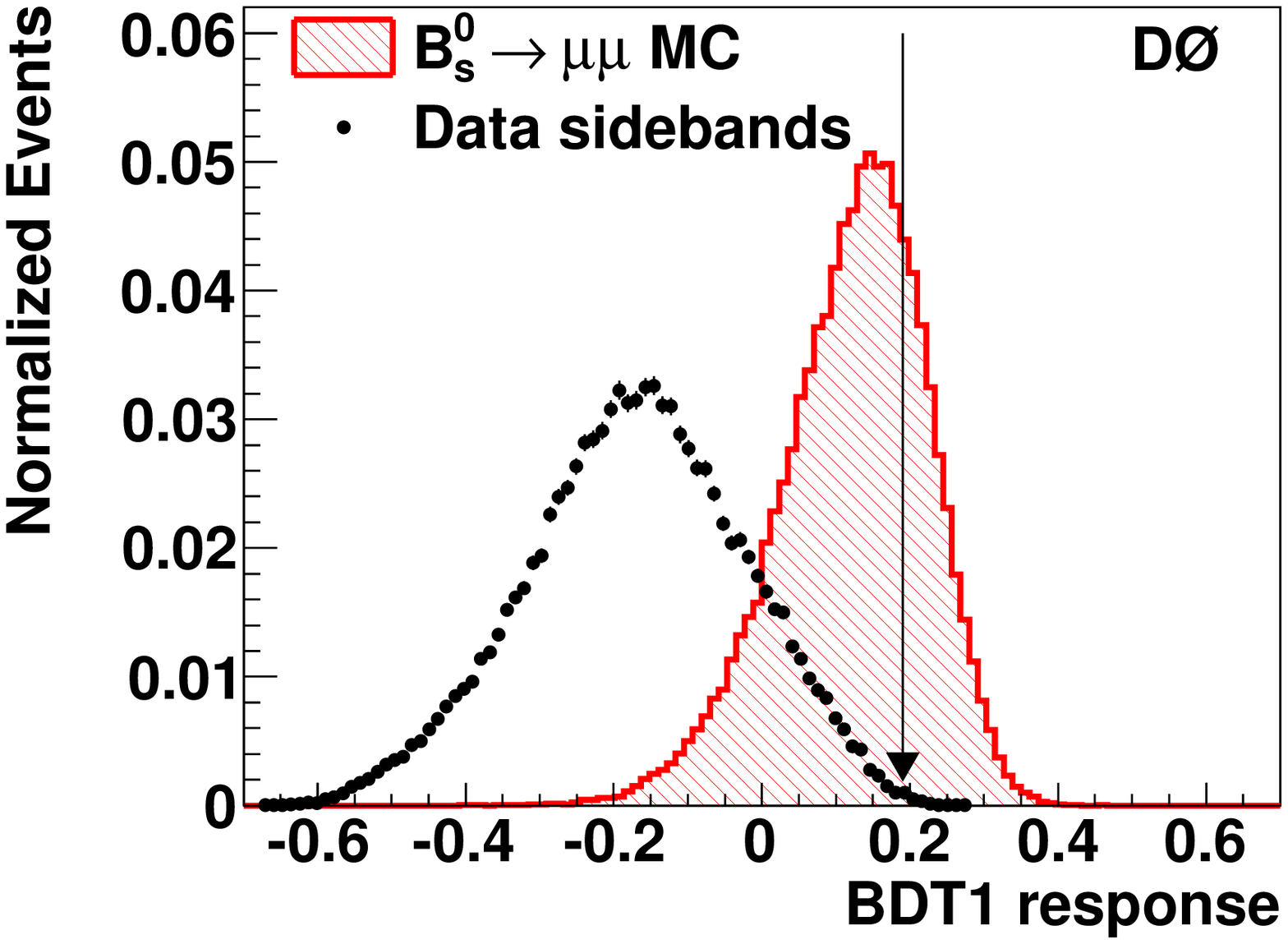}}
\subfigure[]{\label{KS2}\includegraphics [width=3.0in] {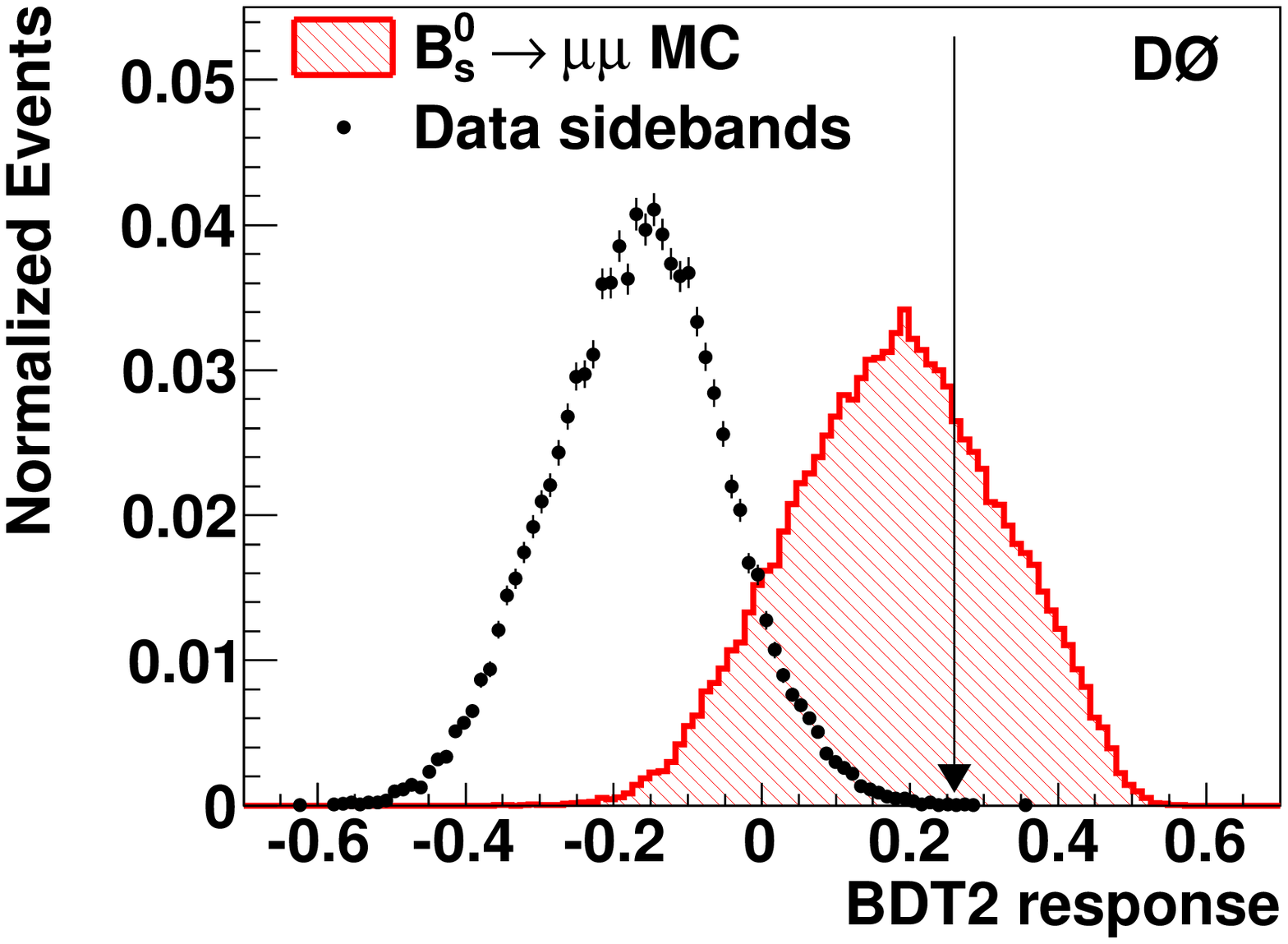}}
\caption{(color online) Distributions of the BDT response for (a) BDT1, trained against 
sequential decay backgrounds, and (b) BDT2, trained against double $B$ decay 
backgrounds. MC simulation is used for the signal, while the data sidebands are used for 
the backgrounds. The  vertical lines denote the BDT selection cuts in the 
analysis. All distributions are normalized to unit area. }
\label{KS}
\end{center}
\end{figure*}

To improve the statistics available for training, the data 
epochs 
are combined and used together to train the BDT.  The signal MC samples for 
each data epoch are combined according to the integrated luminosity for each 
epoch into a common sample. The data sidebands and signal MC are then 
randomly split into three samples.  Sample A, with 25\% of the events, is used 
to train the BDTs.  Sample B, with 25\% of the events, is used to optimize the 
selections on the BDT response.  Sample C, with 50\% of the events, is 
used to 
determine the expected signal (from the MC sample) and background (from the 
data sideband sample) yields. 
The results of the 
{\sc TMVA} BDT training for both BDT1, trained to remove sequential decay 
backgrounds, 
and BDT2, trained to remove double semileptonic $B$ meson decays, can be 
seen in Fig.~\ref{KS}.  We  check that the response of both BDT 
discriminants is independent of dimuon mass over the relevant mass range. 
The optimal BDT selections are determined by optimizing the 
expected limit on  ${\cal B}(B_s^0 \to \mu^+ \mu^-)$ and are found to be 
BDT1 $>0.19$ and BDT2 $>0.26$.  

\section {Background estimates and expected limit}  
\begin {figure*} [!th] 
\begin{center} 
\subfigure[]{\label{bdtcutsa}\includegraphics [width=3.0in] {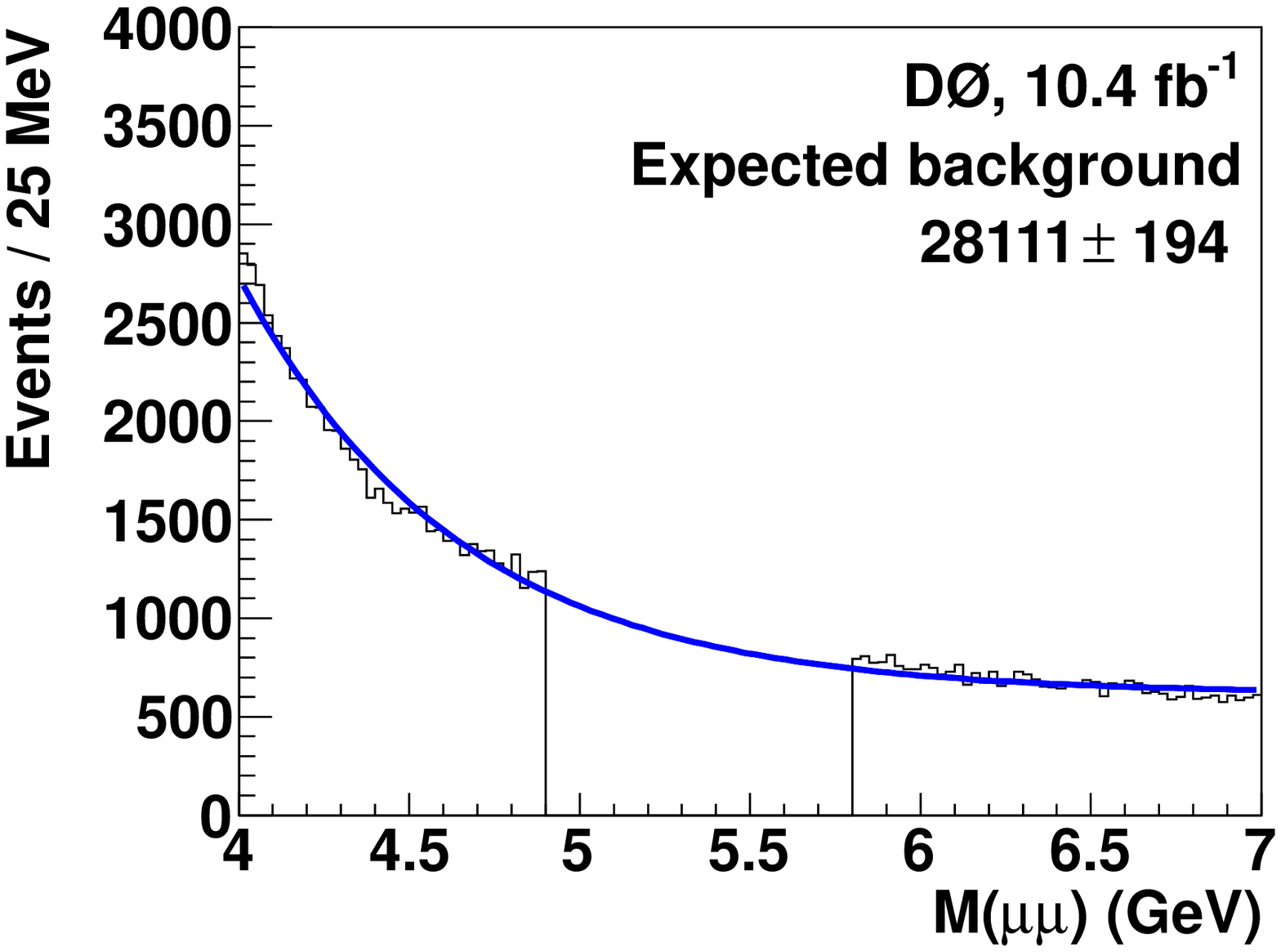}} 
\subfigure[]{\label{bdtcutsb}\includegraphics [width=3.0in] {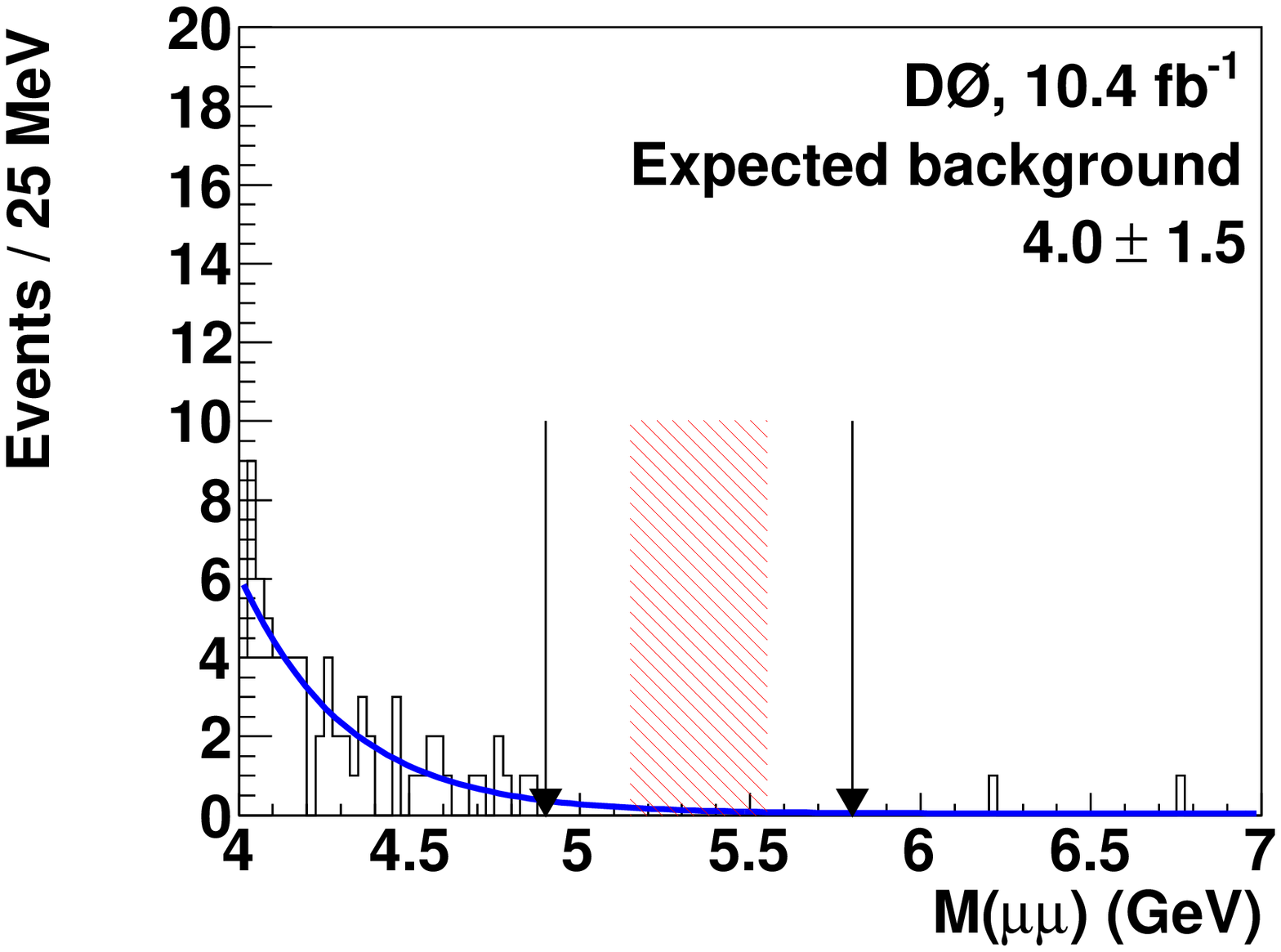}}
\caption{(color online) Dimuon mass 
distribution for sample C (a) before and (b) after BDT selection 
cuts.  The edges of the 
blinded region are denoted in (b) by the vertical lines 
at 4.9 and 5.8 GeV, and the shaded area denotes the signal window. 
The curves are fits to an exponential plus constant function. The numbers 
of expected background events are determined from an interpolation of the fit 
into the signal window and scaled to the full dataset.} 
\label{bdtcuts} 
\end{center} 
\end{figure*}

 Figure~\ref{bdtcuts} shows the blinded dimuon mass distributions
 before (Fig.~\ref{bdtcutsa}) and after (Fig.~\ref{bdtcutsb}) 
the BDT selection cuts 
for the half of the data (sample C) used to estimate the 
number of background events. The signal window within the 
blinded region is chosen to 
maximize the signal significance $S/\sqrt{S+B}$, where $S$ is the expected 
number of signal events as determined from the SM branching fraction, and 
$B$ is the expected background. The 
number of expected background events is determined by a likelihood fit to 
the data in the sideband
regions, which is then interpolated into the blinded region. The optimum signal 
region is determined to be $\pm 1.6 \sigma$ centered on the $B_s^0$ mass, where 
$\sigma$ = 125~MeV is the average width 
of the double Gaussian used to fit the 
dimuon mass distribution in the $B_s^0 \to \mu^+ \mu^-$ MC sample.  
The blinded region includes a control region of width $2\sigma$ on 
each side of the signal window. 
While only half of the dataset is shown, the numbers of expected 
background 
events quoted in Fig.~\ref{bdtcuts} are scaled to the full dataset. The 
numbers given are for the 
estimated dimuon background events in the signal region. 
  
The efficiency for retaining signal events when all BDT selections are applied,
including the pre-training cuts (see Sec.~\ref{bdt}) and the final BDT cuts,
is determined to be 0.12 $\pm$ 0.01, where the error is due to 
variation over the different data epochs. We obtain a final SES of 
(2.8 $\pm$ 0.24)$\times 10^{-9}$, corresponding to an expected number of 
signal events at the SM branching fraction of 1.23 $\pm$ 0.13.
 For the 
dimuon background the expected number of events in the signal and control 
regions is 
determined by applying a log likelihood fit to the dimuon mass 
distribution using an exponential plus constant functional form.
 The fit is performed excluding the blinded region, and 
the resulting fit is interpolated into the signal and control regions. 
This procedure yields an expected number of dimuon background events 
in the signal region of 4.0 $\pm$ 1.5 events, where the uncertainty is 
only statistical. The corresponsing estimate for the 
expected number of events in the control region is $6.7 \pm 2.6$ events, 
with $5.3 \pm 1.9$ events expected in the lower control region (dimuon 
masses from 4.9 to 5.15~GeV), and $1.4 \pm 1.4$ events in the upper control 
region (dimuon masses from 5.55 to 5.8~GeV). 
To determine the systematic uncertainty on the 
background estimate, we use other functional forms for the background fit, 
resulting in a systematic uncertainty of 0.6 events. Adding the statistical 
and systematic errors in quadrature yields a final dimuon background 
estimate in the signal region of 4.0 $\pm$ 1.6 events and 
$6.7 \pm 2.7$ events in the control region. 

In addition to the dimuon background, there is background from 
the decay mode $B_s^0 \to 
K^+K^-$, which has kinematics very similar to the signal. We estimate this 
background by scaling the expected number of signal events by the 
appropriate branching fractions \cite{pdg12} and by the ratio of the 
probabilities for both $K$ mesons to be misidentified as muons, 
$\epsilon(KK\to \mu \mu)$, to 
the probability that two muons are correctly identified as muons, 
$\epsilon(\mu \mu \to \mu \mu)$. The probability 
that a $K$ meson is misidentified as a muon is measured 
in the data using $D^0 \to K \pi$ decays. We assume that the 
probability of two $K$ mesons being misidentified as muons is the 
product of the 
probabilities for each individual $K$ meson.  The muon identification efficiency 
is measured in the data from $J/\psi \to \mu \mu$ decays.  The efficiency 
ratio $\epsilon(KK\to \mu \mu)/\epsilon(\mu \mu \to \mu \mu)$ is determined 
to be $(3.0 \pm 1.1) \times 10^{-5}$. We estimate the background from 
$B_s^0 \to KK$ decays to be 0.28 $\pm$ 0.11 events. We also find a 
consistent 
estimate of this background using a $B_s^0 \to KK$ MC sample. Other 
possible peaking backgrounds such as $B^0_d \to K\pi$ and $B_s^0 \to K\pi$ 
are negligible due to the combination of smaller branching fractions and a 
$\pi \to \mu$ misidentification probability that is more than a factor of 
10 smaller than the $K \to \mu$ misidentification probability in the D0 
detector.

We set an upper limit on the $B_s^0 \to \mu^{+} \mu^{-}$ branching 
fraction
using the CL$_s$, or modified frequentist method \cite{junk}.  A Poisson 
likelihood function is used to 
calculate the number of signal events which would occur with a probability of 
0.05 (for a 95\% CL upper  confidence limit) when $N_{\text{obs}}$ data events are 
observed in the signal region with a known expected number of background events.

The limit calculation includes a convolution over probability 
distributions representing the uncertainties in the background and the 
signal.  The uncertainty in the $B_s^0 \to KK$ peaking background 
is assumed to be Gaussian. The dimuon background in the 
signal region is estimated by the fit shown in Fig.\ \ref{bdtcutsb}. The 
normalized likelihood function from this fit is used as the probability 
distribution function for the dimuon background in the convolution.  
The expected 
number of signal events, assuming the SM branching fraction, is 1.23 $\pm$ 0.13 
events, with the uncertainty assumed to be Gaussian. The total 
expected background is 4.3 $\pm$ 1.6 events. 
 Weighting each possible outcome by its Poisson probability 
yields an expected 95\% C.L. upper limit on the branching fraction ${\cal 
B}(B_s^0 
\to \mu^+\mu^-)$
of $23 \times 10^{-9}$. 

\begin {figure}[h!] 
\begin{center} 
\includegraphics [width=3.5in] {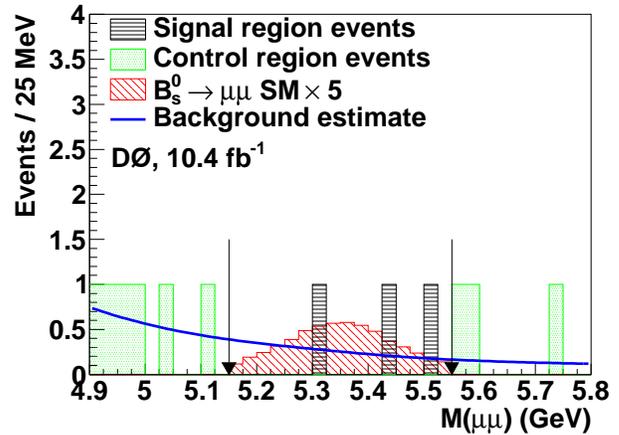} 
\caption{(color online) Dimuon mass distribution in the blinded region 
for the full dataset after BDT selections are applied.  
The curve shows  the fit from Fig.\ 
\ref{bdtcutsb} used to 
determine the expected number of background events.
The SM expectation for signal events multiplied by five is also indicated.  
The vertical lines mark the edge of the signal window.} 
\label{final_mass} 
\end{center} 
\end{figure} 

\begin {figure}[h!] 
\begin{center} 
\includegraphics [width=3.5in] {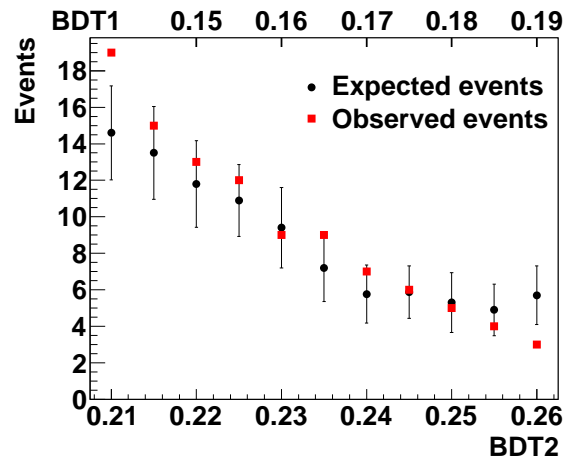} 
\caption{(color online) Expected number of events and observed number of 
events in the 
 signal region as the two BDT cuts are relaxed in parallel.  The 
expected 
number of events includes the 
 dimuon background, the $B_s^0 \to KK$ background, and the expected number of 
signal events. The upper horizontal axis shows the cut applied to BDT1, 
while the lower horizontal axis shows the cut applied to BDT2.
}
\label{marjplot} 
\end{center} 
\end{figure} 
Upon unblinding, a total of nine events is found in the control region 
above and below the signal region, as shown in Fig.\ \ref{final_mass}. 
Six events are found in the control region below the signal window, and 
three events
are found in the control region above the signal window.  
This number of events and their distribution within the control regions 
is in agreement 
with the expected number of background events interpolated from the data 
sidebands. As seen in Fig.\ \ref{final_mass}, three 
events are found in the dimuon mass 
signal window, in agreement with the expected background and also 
with the expected signal + background. 
We check that the properties of all events 
found in the blinded region, such as the $p_T$ of the dimuon system, the 
$p_T$ of the individual muons, the dimuon pointing angle, and the various isolation 
quantities, are consistent with expectations. 
We also check that, as the BDT cuts are relaxed, the 
number of events observed in the signal region remains in good 
agreement with expectations, as shown in Fig.\ \ref{marjplot}.

The observed number of events and the SES allow us to set a 95\% C.L. 
upper limit  $\cal{B}$$(B_s^0 \to \mu^{+} \mu^{-}) < 15 \times 10^{-9}$.

\section {Summary} In summary, we have searched for the rare decay $B_s^0 
\to \mu^+ \mu^-$ in the full D0 dataset.  We employ two Boosted Decision 
Tree multivariate discriminators, one trained to discriminate against 
sequential decays $b(\bar{b}) \to c\mu^-(\bar{c}\mu^+)X$ followed by 
$c(\bar{c}) \to \mu^+ (\mu^-)X$ and the other to discriminate against 
double semileptonic decays $b\to \mu^- X$ and $\bar{b} \to \mu^+ X$.  
The sidebands around the signal region in the dimuon invariant mass distribution 
are used to 
estimate the dominant backgrounds. The 
expected limit is 23 $\times 10^{-9}$, and the expected background (signal) in the 
signal region is 4.3 $\pm$ 1.6 (1.23 $\pm$ 0.13) events.  We observe three events in the 
signal region consistent with expected background. The probability that 
the background alone (signal + background) could produce the observed number of events or a 
larger number of events in the signal region is 0.77 (0.88).  We set an observed 
95\% C.L. upper limit $\cal{B}$$(B_s^0 \to \mu^{+} \mu^{-}) < 15 \times 
10^{-9}$. This upper limit supersedes the previous D0 95\% C.L. limit of 
51 $\times 10^{-9}$ \cite {masato}, and improves upon that limit by a 
factor of 3.4. The improvement in the expected limit is a factor of 1.7 
greater than the improvement that would be expected due to increased 
luminosity alone. The additional 
improvement arises from the inclusion of several isolation-type variables 
in the multivariate discriminants and in the use of two separate 
discriminants to distinguish backgrounds from sequential $b$ 
quark decays and double $b$ quark decays. This result is the most 
stringent Tevatron limit and is 
compatible with the recent evidence of this decay produced by the LHCb
experiment \cite{newlhcb}.

We thank the staffs at Fermilab and collaborating institutions,
and acknowledge support from the
DOE and NSF (USA);   
CEA and CNRS/IN2P3 (France);
MON, NRC KI and RFBR (Russia);
CNPq, FAPERJ, FAPESP and FUNDUNESP (Brazil);
DAE and DST (India);
Colciencias (Colombia);
CONACyT (Mexico);
NRF (Korea);
FOM (The Netherlands);
STFC and the Royal Society (United Kingdom);
MSMT and GACR (Czech Republic);
BMBF and DFG (Germany);
SFI (Ireland);
The Swedish Research Council (Sweden);
and
CAS and CNSF (China).
%
%\vspace*{.3in}

\end{document}

%% file: author_list.tex
%
\affiliation{LAFEX, Centro Brasileiro de Pesquisas F\'{i}sicas, Rio de Janeiro, Brazil}
\affiliation{Universidade do Estado do Rio de Janeiro, Rio de Janeiro, Brazil}
\affiliation{Universidade Federal do ABC, Santo Andr\'e, Brazil}
\affiliation{University of Science and Technology of China, Hefei, People's Republic of China}
\affiliation{Universidad de los Andes, Bogot\'a, Colombia}
\affiliation{Charles University, Faculty of Mathematics and Physics, Center for Particle Physics, Prague, Czech Republic}
\affiliation{Czech Technical University in Prague, Prague, Czech Republic}
\affiliation{Center for Particle Physics, Institute of Physics, Academy of Sciences of the Czech Republic, Prague, Czech Republic}
\affiliation{Universidad San Francisco de Quito, Quito, Ecuador}
\affiliation{LPC, Universit\'e Blaise Pascal, CNRS/IN2P3, Clermont, France}
\affiliation{LPSC, Universit\'e Joseph Fourier Grenoble 1, CNRS/IN2P3, Institut National Polytechnique de Grenoble, Grenoble, France}
\affiliation{CPPM, Aix-Marseille Universit\'e, CNRS/IN2P3, Marseille, France}
\affiliation{LAL, Universit\'e Paris-Sud, CNRS/IN2P3, Orsay, France}
\affiliation{LPNHE, Universit\'es Paris VI and VII, CNRS/IN2P3, Paris, France}
\affiliation{CEA, Irfu, SPP, Saclay, France}
\affiliation{IPHC, Universit\'e de Strasbourg, CNRS/IN2P3, Strasbourg, France}
\affiliation{IPNL, Universit\'e Lyon 1, CNRS/IN2P3, Villeurbanne, France and Universit\'e de Lyon, Lyon, France}
\affiliation{III. Physikalisches Institut A, RWTH Aachen University, Aachen, Germany}
\affiliation{Physikalisches Institut, Universit\"at Freiburg, Freiburg, Germany}
\affiliation{II. Physikalisches Institut, Georg-August-Universit\"at G\"ottingen, G\"ottingen, Germany}
\affiliation{Institut f\"ur Physik, Universit\"at Mainz, Mainz, Germany}
\affiliation{Ludwig-Maximilians-Universit\"at M\"unchen, M\"unchen, Germany}
\affiliation{Fachbereich Physik, Bergische Universit\"at Wuppertal, Wuppertal, Germany}
\affiliation{Panjab University, Chandigarh, India}
\affiliation{Delhi University, Delhi, India}
\affiliation{Tata Institute of Fundamental Research, Mumbai, India}
\affiliation{University College Dublin, Dublin, Ireland}
\affiliation{Korea Detector Laboratory, Korea University, Seoul, Korea}
\affiliation{CINVESTAV, Mexico City, Mexico}
\affiliation{Nikhef, Science Park, Amsterdam, the Netherlands}
\affiliation{Radboud University Nijmegen, Nijmegen, the Netherlands}
\affiliation{Joint Institute for Nuclear Research, Dubna, Russia}
\affiliation{Institute for Theoretical and Experimental Physics, Moscow, Russia}
\affiliation{Moscow State University, Moscow, Russia}
\affiliation{Institute for High Energy Physics, Protvino, Russia}
\affiliation{Petersburg Nuclear Physics Institute, St. Petersburg, Russia}
\affiliation{Instituci\'{o} Catalana de Recerca i Estudis Avan\c{c}ats (ICREA) and Institut de F\'{i}sica d'Altes Energies (IFAE), Barcelona, Spain}
\affiliation{Uppsala University, Uppsala, Sweden}
\affiliation{Lancaster University, Lancaster LA1 4YB, United Kingdom}
\affiliation{Imperial College London, London SW7 2AZ, United Kingdom}
\affiliation{The University of Manchester, Manchester M13 9PL, United Kingdom}
\affiliation{University of Arizona, Tucson, Arizona 85721, USA}
\affiliation{University of California Riverside, Riverside, California 92521, USA}
\affiliation{Florida State University, Tallahassee, Florida 32306, USA}
\affiliation{Fermi National Accelerator Laboratory, Batavia, Illinois 60510, USA}
\affiliation{University of Illinois at Chicago, Chicago, Illinois 60607, USA}
\affiliation{Northern Illinois University, DeKalb, Illinois 60115, USA}
\affiliation{Northwestern University, Evanston, Illinois 60208, USA}
\affiliation{Indiana University, Bloomington, Indiana 47405, USA}
\affiliation{Purdue University Calumet, Hammond, Indiana 46323, USA}
\affiliation{University of Notre Dame, Notre Dame, Indiana 46556, USA}
\affiliation{Iowa State University, Ames, Iowa 50011, USA}
\affiliation{University of Kansas, Lawrence, Kansas 66045, USA}
\affiliation{Louisiana Tech University, Ruston, Louisiana 71272, USA}
\affiliation{Northeastern University, Boston, Massachusetts 02115, USA}
\affiliation{University of Michigan, Ann Arbor, Michigan 48109, USA}
\affiliation{Michigan State University, East Lansing, Michigan 48824, USA}
\affiliation{University of Mississippi, University, Mississippi 38677, USA}
\affiliation{University of Nebraska, Lincoln, Nebraska 68588, USA}
\affiliation{Rutgers University, Piscataway, New Jersey 08855, USA}
\affiliation{Princeton University, Princeton, New Jersey 08544, USA}
\affiliation{State University of New York, Buffalo, New York 14260, USA}
\affiliation{University of Rochester, Rochester, New York 14627, USA}
\affiliation{State University of New York, Stony Brook, New York 11794, USA}
\affiliation{Brookhaven National Laboratory, Upton, New York 11973, USA}
\affiliation{Langston University, Langston, Oklahoma 73050, USA}
\affiliation{University of Oklahoma, Norman, Oklahoma 73019, USA}
\affiliation{Oklahoma State University, Stillwater, Oklahoma 74078, USA}
\affiliation{Brown University, Providence, Rhode Island 02912, USA}
\affiliation{University of Texas, Arlington, Texas 76019, USA}
\affiliation{Southern Methodist University, Dallas, Texas 75275, USA}
\affiliation{Rice University, Houston, Texas 77005, USA}
\affiliation{University of Virginia, Charlottesville, Virginia 22904, USA}
\affiliation{University of Washington, Seattle, Washington 98195, USA}
\author{V.M.~Abazov} \affiliation{Joint Institute for Nuclear Research, Dubna, Russia}
\author{B.~Abbott} \affiliation{University of Oklahoma, Norman, Oklahoma 73019, USA}
\author{B.S.~Acharya} \affiliation{Tata Institute of Fundamental Research, Mumbai, India}
\author{M.~Adams} \affiliation{University of Illinois at Chicago, Chicago, Illinois 60607, USA}
\author{T.~Adams} \affiliation{Florida State University, Tallahassee, Florida 32306, USA}
\author{G.D.~Alexeev} \affiliation{Joint Institute for Nuclear Research, Dubna, Russia}
\author{G.~Alkhazov} \affiliation{Petersburg Nuclear Physics Institute, St. Petersburg, Russia}
\author{A.~Alton$^{a}$} \affiliation{University of Michigan, Ann Arbor, Michigan 48109, USA}
\author{A.~Askew} \affiliation{Florida State University, Tallahassee, Florida 32306, USA}
\author{S.~Atkins} \affiliation{Louisiana Tech University, Ruston, Louisiana 71272, USA}
% ...
%
% visitor_addresses.tex                       11 January 2013 
%  available symbols are:
%  $\ast, \dag, \ddag, \S, \P, $\|$, $\ast\ast$, \dag\dag, \ddag\ddag ,\#
%
\collaboration{The D0 Collaboration\footnote{with visitors from
%{alton}
$^{a}$Augustana College, Sioux Falls, SD, USA,
%{burdin}
$^{b}$The University of Liverpool, Liverpool, UK,
%{garcia-guerra}
$^{c}$UPIITA-IPN, Mexico City, Mexico,
%{grohsjean}
$^{d}$DESY, Hamburg, Germany,
%{partridge}
$^{e}$SLAC, Menlo Park, CA, USA,
%{hesketh}
$^{f}$University College London, London, UK,
%{luna-garcia}
$^{g}$Centro de Investigacion en Computacion - IPN, Mexico City, Mexico,
%{podesta-lerma}
$^{h}$ECFM, Universidad Autonoma de Sinaloa, Culiac\'an, Mexico,
%{santos}
$^{i}$Universidade Estadual Paulista, S\~ao Paulo, Brazil,
%{meyer}
$^{j}$Karlsruher Institut f\"ur Technologie (KIT) - Steinbuch Centre for Computing (SCC)
and
%{patwa}
$^{k}$Office of Science, U.S. Department of Energy, Washington, D.C. 20585, USA.
%{falkowski}
%$^{?}$Laboratoire de Physique Theorique, Orsay, FR,
%{hooper}
%$^{?}$Visitor from Bradley University, Peoria, IL, USA.
%{kozminski}
%$^{?}$}Visitor from Lewis University, Romeoville, IL, USA.
%{weber}
%$^{?}$Universit{\"a}t Bern, Bern, Switzerland.
%{deceased}
%$^{\ddag}$Deceased.
}} \noaffiliation
\vskip 0.25cm

%% file: bspaper_v4_5_sub.bbl
\begin{thebibliography} {99}
\bibitem {eta} Pseudorapidity $\eta$ is defined as 
$\eta=-\text{ln}[\text{tan}(\theta/2)]$, where $\theta$ is the
polar angle measured with respect to the proton beam direction.
\end{thebibliography}
